\documentclass[twocolumn,times,tighten,trackchanges]{aastex701} 

\usepackage{amsmath}
\usepackage{comment}
\usepackage{graphicx}
\usepackage[utf8]{inputenc}
\usepackage{natbib}
\usepackage{subcaption}
\graphicspath{{./}{figures/}}

\newcommand{\ifm}[1]{\relax\ifmmode#1\else$\mathsurround=0pt #1$\fi}
\newcommand{\be}{\begin{equation}}
\newcommand{\ee}{\end{equation}}
\newcommand{\bea}{\begin{eqnarray}}
\newcommand{\eea}{\end{eqnarray}}

\newcommand{\sggg}[1]{\textcolor{green}{[]}}

\def\m{{\rm\thinspace m}}

\def\pc{{\rm\thinspace pc}}
\def\kpc{{\rm\thinspace kpc}}

\def\Mpc{{\rm\thinspace Mpc}}

\def\g{{\rm\thinspace g}}

\def\Msun{\hbox{$\rm\thinspace M_{\odot}$}}

\def\s{{\rm\thinspace s}}       
\def\yr{{\rm\thinspace yr}}

\def\Gyr{{\rm\thinspace Gyr}}

\def\Msunpc2{{\Msun\pc}^{-2}}
\def\Msunyrkpc2{{\Msun\yr^{-1}\kpc}^{-2}}

\def\magarcsec2{{\rm\thinspace mag\thinspace arcsec}^{-2}}

\shorttitle{Multiple Galaxy Mergers in $\Lambda$CDM}
\shortauthors{Mack, J; Genel, S}

\begin{document}

\title{On the Frequency of Multiple Galaxy Mergers in $\Lambda$CDM Cosmological Simulations}
\author{Jonathan Mack}
\affiliation{Independent researcher}
\email{jkmack@proton.me}
\author[0000-0002-3185-1540]{Shy Genel}
\affiliation{Center for Computational Astrophysics, Flatiron Institute, 162 5th Avenue, New York, NY, 10010, USA}
\affiliation{Columbia Astrophysics Laboratory, Columbia University, 550 West 120th Street, New York, NY, 10027, US}
\email{sgenel@flatironinstitute.org}

\begin{abstract}
Mergers are believed to play a pivotal role in galaxy evolution, and measuring the galaxy merger fraction is a longstanding goal of both observational and theoretical studies. In this work, we extend the consideration of the merger fraction from the standard measure of binary mergers, namely those comprising two merging galaxies, to multiple mergers, namely mergers involving three or more galaxies. We use the Illustris and IllustrisTNG cosmological hydrodynamical simulations to provide a theoretical prediction for the fraction of galaxy systems that are involved in a multiple merger as a function of various parameters, with a focus on the relationship between the multiple merger fraction $f_m$ and the total merger fraction $f_t$. We generally find that binary mergers dominate the total fraction and that $f_m\approx (0.5-0.7)f_t^{5/3}$, a prediction that can be tested observationally. We further compare the empirical simulation results with toy models where mergers occur, on the evolution timeline of a galaxy, either at constant intervals or as a Poisson process at a constant rate. From these comparisons, where the toy models typically produce lower multiple merger fractions, we conclude that in cosmological simulations, mergers are more strongly clustered in time than in these toy scenarios, likely reflecting the hierarchical nature of cosmological structure formation.
\end{abstract}

\section{Introduction} \label{sec:intro}

Considerable effort has been dedicated over the decades to elucidating the mechanisms that contribute to the evolution of galaxies over time, with galaxy-galaxy mergers constituting a particularly attractive one. Individual systems of merging galaxies in various stages, which show galaxies being distorted in diverse ways, provide direct evidence that mergers can serve a transformational role on the evolution timeline of a galaxy. Aggregate statistics, such as the galaxy-galaxy merger rate and fraction, hold at least as important clues. Observational studies (e.g.~\citealp{1996ApJS..107....1A,2009A&A...498..379D,BluckA_12a,ManA_16a}) have largely concluded that the merger rate increases toward higher redshifts, although the precise shape of this dependence remains subject to considerable uncertainty (e.g.~\citealp{2019A&A...631A..87V}).

Mergers can be identified observationally in various ways. The most basic classification is between galaxies that are about to merge, i.e.~galaxy pairs (e.g.~\citealp{ManthaK_18a}), and galaxies that are recent merger remnants, which are identified based on their morphologies, kinematics, and/or the presence of strong tidal features such as tails and shells around them (e.g.~\citealp{BridgeC_10a,LacknerL_14a}). One of the challenges in quantifying galaxy mergers in meaningful ways is that different merger identification methods produce different merger populations and result in different aggregate merger statistics. One specific challenge among many is the difficulty of estimating the duration of the observable merger phase. Another is the difficulty of estimating the mass ratio of the merger, which is considered a key parameter distinguishing for example between the rarer `major' (roughly equal mass) and the more frequent but less consequential `minor' mergers.

Galaxy mergers are usually considered to occur between two progenitor galaxies; however, multiple merger systems, i.e.~those involving three or more galaxies, have also been observed and studied in detail (e.g.~\citealp{Grajales_Medina_2022}), with Stephen's Quintet \citep{DucP_18a} and the Leo triplet \citep{Wu_2022} being rather well-known examples. Working with larger samples, \citet{OMillA_12a} compiled a catalog of 1,092 isolated triplets of galaxies, some of which are actively merging, while \citet{Darg_2011} determined using the Galaxy Zoo \citep{LintottC_08a} that the multiple merger fraction is at least two orders of magnitude smaller than the binary merger fraction. In the high redshift Universe, James Web Space Telescope observations found extremely compact configurations of multiple galaxies that likely represent imminent multiple mergers (e.g.~\citealp{Wylezalek_2022,hashimoto2023rioja}). However, there is no consensus on the prevalence of multiple mergers and relatively few studies on their implications for galaxy formation.

Multiplicity in mergers is routinely identified not only in mergers of individual galaxies but also in mergers of galaxy groups and clusters. For example, \citet{Tempel_2017} found using Sloan Digital Sky Survey \citep{YorkD_00a} data that 8\% of the galaxy groups systems that are undergoing a merger contained three or more merging subgroups. Some specific systems of this kind have been analyzed in detail \citep{Golovich_2016,van_Weeren_2017,Ruggiero_2018,Schellenberger_2019,Sohn_2019,Sarkar_2023,jimenezteja2023dissecting,2023arXiv230901716R}, generally finding complex dynamics.

Beginning in the 1970s, computer simulations became important tools in investigating galaxy evolution, and some of the very first numerical results clearly indicated that galaxy mergers can play a pivotal role in the evolution of a galaxy and its properties. The main initial work in this area was done by \citet{1972ApJ...178..623T}, who used gravity-only simulations to show that observed galaxy bridges and tails are relics of close encounters with other galaxies. Subsequent simulations, ever larger and increasingly commonly including hydrodynamics, gave rise to the now-classical picture of two merging disk galaxies producing a spheroidal remnant \citep{1992ApJ...400..460H}, demonstrated the buildup of a strong central mass concentration during mergers due to torques and gas dissipation \citep{HernquistL_95a} and the resulting triggering of Active Galactic Nuclei \citep{SpringelV_05d}, and suggested many other transformational effects of mergers on the participating galaxies (e.g.~\citealp{2003ApJ...597..893N,SpringelV_05b,HopkinsP_07a,RenaudF_15a}).

The advent of large-scale cosmological N-body simulations that can resolve the host halos of individual galaxies, including their subhalo populations, led to the ability to make $\Lambda$CDM-based merger rate predictions. \citet{2008MNRAS.386..577F} and \citet{2009ApJ...701.2002G} constructed merger trees from the Millennium simulation \citep{SpringelV_05a} and determined the halo merger rate as a function of descendant halo mass, progenitor mass ratio, and redshift. The main findings included that the merger rate increases with redshift and mass, as well as with mass ratio, so that minor mergers are more numerous than major ones. \citet{2009ApJ...702.1005S} quantified the halo merger rates for both infall and destruction. Further studies populated cosmological N-body simulations with galaxies using either Semi-Analytical Models (SAM; \citealp{GuoQ_07a}) or semi-empirical models \citep{Wetzel_2009,2010ApJ...715..202H} to predict galaxy-galaxy merger rates, and found meaningful differences between the merger rates of halos and those of galaxies (see also \citealp{HopkinsP_10a}). All of these studies relied on the construction of merger trees from the simulations; significant work has been devoted to the development of merger tree construction algorithms and comparisons between them to provide more accurate and robust estimates of the merger rate (e.g.~\citealp{2009A&A...506..647T,2012ascl.soft10008B,2014MNRAS.441.3488A}).

The subsequent development of similarly large cosmological hydrodynamical simulations that explicitly model the formation of galaxies enabled measuring galaxy-galaxy merger rates directly, without relying on halos or subhalos as an intermediate step, and studying their various effects on large galaxy populations. For example, the Illustris simulation \citep{2014Natur.509..177V,2014MNRAS.444.1518V,GenelS_14a}, a $100\Mpc/h$ cosmological box evolved with moving mesh hydrodynamics, was used by \citet{2015MNRAS.449...49R} to perform a galaxy-galaxy merger rate measurement, by \citet{Rodriguez_Gomez_2016} to investigate the merger contribution to galaxy mass, and by \citet{2017MNRAS.467.3083R} to study the effect of mergers on galaxy morphology.

While merger tree analysis of cosmological simulations inherently provides the galaxy merger rate, defined (for any given choice of parameters such as mass ratio) as the number of merger events occurring per unit time, observations are only able to directly probe the merger fraction, defined as the fraction of galaxy systems that at a given time are in the process of undergoing a merger (again, for any given choice of selection parameters and merger properties). The relation between these two quantities, and therefore between theoretical and observational constraints on galaxy mergers, goes through the concept of the merger duration, namely the length of the time window during which a merger is observable as such \citep{2011ApJ...742..103L}. Estimates of merger durations and their dependence on various parameters have been made using multiple hydrodynamic merger simulations and cosmological simulations \citep{LotzJ_08b,HungC_16a,2017MNRAS.468..207S}. The concept of the merger duration, as we show in this paper, is also crucial for the quantification of the multiple merger fraction.

Theoretical research on multiple mergers is relatively limited, but some numerical investigations have yielded insights in related domains. For example, using the Millennium simulation, \citet{Moreno_2013} showed that only a small fraction (9\%) of galaxy pairs correspond to the idealized isolated case typically studied with merger simulations. \citet{Moster_2013} simulated sequences of mergers of idealized halos and galaxies that are motivated by SAM merger trees, which include multiple mergers in proximity. \citet{An_2019} used simulations to determine that at $z=0$ flybys are more common than mergers for both binary and multiple interactions. \citet{D_az_Gim_nez_2019} investigated the effect of varying SAM parameters, including cosmological parameters, on the nature and frequency of compact groups of galaxies, a class of objects closely associated with multiple mergers. Most recently, \citet{Ni_2022} showed that an ultra-massive black hole (UMBH, $M_{\rm BH} > 10^{10} M_{\odot}$) in the ASTRID cosmological hydrodynamical simulation was formed from a successive (possibly multiple) galaxy merger. Finally, \citet{Darg_2011} compared the multiple merger fraction in the Millennium simulation with that of observed galaxies from the Galaxy Zoo and found substantial agreement.

Despite the above, there has been little work on quantifying the frequency of multiple mergers in large-scale cosmological hydrodynamical simulations, which is the main goal of this paper. We organize the paper as follows. In Section \ref{sec:methods} we describe our methods, in Section \ref{sec:results} we present our results, and in Section \ref{sec:conclusion} we summarize the main findings and discuss their significance and potential for future work.

\section{Methods} \label{sec:methods}

Our analysis has three major components. After detailing how we identify mergers (Section \ref{subsec:M_merger_ID}), we define and calculate various types of galaxy merger fractions (Section \ref{subsec:M_f}), and then analyze the time until the closest merger(s) for each galaxy and its relation to the merger fraction (Section \ref{subsec:M_dt}). This is followed by a description of simple toy models for the distribution of time-to-closest-mergers (Section \ref{subsec:M_fits}). Finally, we detail our parameter space of various sample selections, merger definitions, and modeling choices in Section \ref{subsec:M_param}.

\subsection{Merger Identification} \label{subsec:M_merger_ID}

Our primary goal in this work is to investigate the multiple merger fraction, namely the fraction of galaxies at a given time that undergo concurrent mergers with at least two other galaxies. To this end, we use the existing merger trees created with the SubLink algorithm (\citealp{2015MNRAS.449...49R}; hereafter RG15) from the Illustris \citep{2014Natur.509..177V,2014MNRAS.444.1518V,GenelS_14a,SijackiD_14a,NelsonD_15b} and IllustrisTNG simulations \citep{Pillepich_2017a,Springel_2017,Naiman_2018,Marinacci_2018,Nelson_2017,NelsonD_19b}. These trees are constructed with SUBFIND \citep{SpringelV_01} objects, which can be thought of as individual galaxies, by identifying common particles between objects across adjacent snapshots. More specifically, a `direct descendant' is found for each object by identifying the object in a subsequent snapshot that contains most of its particles, and all objects that share their direct descendant are considered to be its `direct progenitors'. A merger appears in the tree whenever an object has more than one direct progenitor, which occurs when SUBFIND stops distinguishing two distinct density peaks within the combined object. Among those direct progenitors, one is defined as the `first progenitor' according to a weighting scheme that most commonly selects the most massive progenitor as such (for details, see RG15), and the others are referred to as `next progenitors' \citep{SpringelV_05a}. We include two versions of SubLink, one that is based on dark matter particles and one that considers baryonic particles.

We postprocess these trees while accounting for a novel concept in merger tree analysis, namely that mergers are not instantaneous but rather have a finite duration. Fundamentally, we must identify which galaxies should be considered as undergoing mergers at any given time. To do so, we identify prospective mergers and analyze their time coverage (Section \ref{subsubsec:M_mrgrID_time}) as we traverse the merger tree (Section \ref{subsubsec:M_mrgrID_traversal}). Further, we address the special case of subhalo "skipping" (Section \ref{subsubsec:M_mrgrID_skip}), and finally discuss the calculation of the merger mass ratio (Section \ref{subsubsec:M_mrgrID_M}).

\subsubsection{Mergers as non-instantaneous events and merger overlap probabilities} \label{subsubsec:M_mrgrID_time}

\begin{figure}[ht!]
\centering
\includegraphics[scale=0.8]{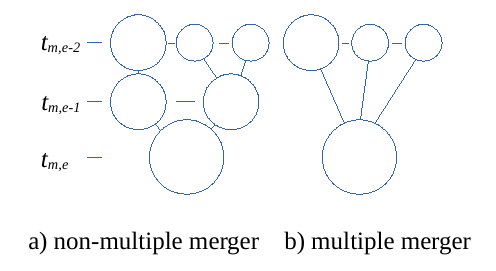}
\caption{A demonstration of the dependence on the snapshot separation in a na{\"i}ve approach to determining whether a merger is a multiple merger, i.e.~between more than two galaxies at once. On the left, case (a) shows how small snapshot separations delineates the mergers in time into a sequence of two binary mergers, with one descendant at snapshot $t_{m,e-1}$ and the final descendant in snapshot $t_{m,e}$, implying this case would not be counted as a multiple merger. In case (b) on the right, however, a larger difference in time between adjacent snapshots $t_{m,e}$ and $t_{m,e-2}$ implies that the three galaxies in snapshot $t_{m,e-2}$ appear to be merging all together into one descendant at $t_{m,e}$, na{\"i}vely appearing as a multiple merger. This demonstrates that relying on the connectivity of the merger tree for identifying multiple mergers is fraught, since that connectivity significantly depends on the snapshot separation.
\label{fig:binary_multiple_trees}}
\end{figure}

The conventional aggregate metric for mergers is the binary (galaxy-galaxy) merger rate, which is characterized by the number of mergers occurring per unit time. For the purpose of analysis, the necessary inputs are typically only the participating masses and the occurrence time of prospective mergers. However, in order to compute statistics of multiple mergers, one must also take into account the finite duration of the merger events. We can understand one of the main reasons for this by considering a successive merger of three galaxies, demonstrated in Figure \ref{fig:binary_multiple_trees}. One might imagine a na{\"i}ve approach that just considers the number of direct progenitors of a galaxy in a merger tree, in which case the answer to whether or not a multiple merger occurs would depend on the arbitrary factor that is the time separation between snapshots in the tree. A larger snapshot separation (as on the right in Figure \ref{fig:binary_multiple_trees}) would show three galaxies merging concurrently, so would be counted as a multiple merger, while a smaller time step (as shown on the left of Figure \ref{fig:binary_multiple_trees}) might resolve the mergers in time, showing a succession of two binary (non-multiple) mergers.

Our immediate goal is to devise a way to identify which time window and which objects in the merger tree any given merger is spanning that is insensitive to the snapshot separation. To achieve this, we first introduce three conceptual time points. The analysis time $t_a$ is the time for which we are looking to measure the merger fraction. The post-merger time $t_{m, e}$ is the time corresponding to the direct descendant of an arbitrary merger in the tree, namely to any object with two or more direct progenitors (see Figure \ref{fig:binary_multiple_trees}), while the pre-merger time $t_{m, s}$ is the time of the snapshot immediately preceding that of $t_{m, e}$, i.e.~the snapshot of the merger's direct progenitors. As we will see, we need to move up and down a merger tree, i.e.~backward and forward in time, to inspect all prospective mergers; while $t_{m, s}$ and $t_{m, e}$ are quantities associated with individual mergers that are encountered as we move up and down a tree, $t_a$ is a constant in any given analysis.

\begin{figure*}[ht!]
\centering
\includegraphics[scale=0.5]{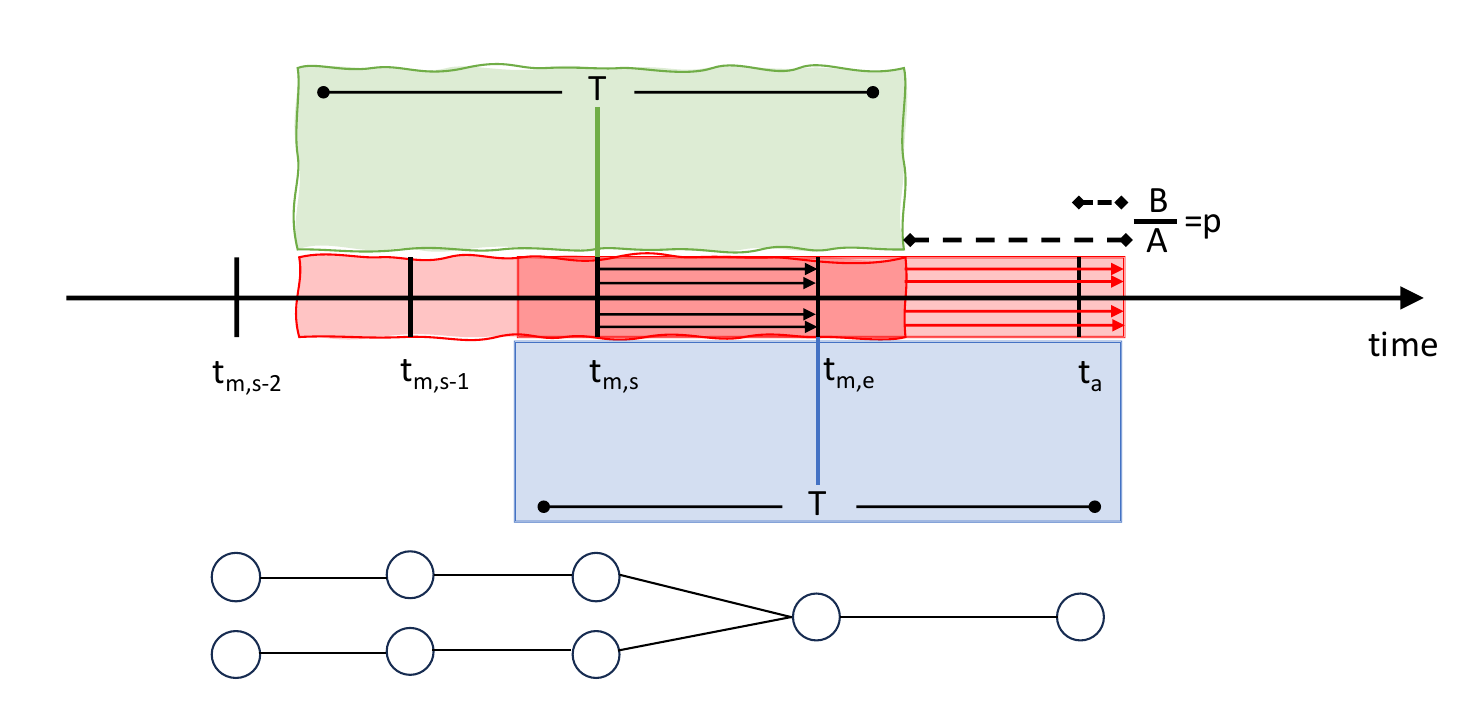}
\caption{An illustration of the role of the concept of the merger duration in our methodology. By identifying a descendant at snapshot $t_{m,e}$ (the post-merger time) and its pair of progenitors at snapshot $t_{m,s}$ (the pre-merger time), we consider the merger to have occurred at some indeterminate time in between (marked by thin black arrows). Since we assume the merger is observable for a time duration $T$ that is centered around the actual merger occurrence time, namely as early as around $t_{m, s}$ (green) or as late as around $t_{m, e}$ (blue), any analysis time point $t_a$ in the range $t_{m, s} - T/2 \le t_a \le t_{m, e} + T/2$ (red) could potentially be a time at which the merger is observable. The probability $p$ that the analysis time $t_a$ is an instant that is actually covered by the $T$-long duration window equals the ratio of two particular time segments, as described by Equation \ref{eq:p} and illustrated here (dashed) for an arbitrarily chosen $t_a$.
\label{fig:overlap}}
\end{figure*}

When mergers are considered as instantaneous events, it is common in the literature (e.g.~RG15) to treat them as having "occurred" at the time of the direct descendant, namely at $t_{m, e}$. However, this notion is arbitrary due to the discreetness of the simulation snapshots, as the descendant of the merger (as per the employed galaxy finder) could have in fact appeared in the simulation at any time between the two adjacent times $t_{m, s}$ and $t_{m, e}$. Furthermore, importantly, mergers are not instantaneous, in some cases taking billions of years to complete. We therefore employ a new term, that of the merger duration $T$, which is the length of the time window during which a merger is `in progress', or observable. We make the assumption here that that time window is symmetric around the true moment of appearance of the descendant, which is anywhere between $t_{m, s}$ and $t_{m, e}$ (but this assumption can be changed, and the results are insensitive to variations around it). Therefore, as depicted in Figure \ref{fig:overlap}, which shows a generic timeline of a merger, the time window during which the merger is potentially observable is bounded by the range $[t_{m, s} - T/2, t_{m, e} + T/2]$. If the analysis time $t_a$ is within this range, i.e.~$t_{m, s} - T/2 \le t_a \le t_{m, e} + T/2$, there is at least some chance that the merger is observable at the analysis time.

The preceding discussion outlines constraints on the time interval during which a given merger is ongoing, but does not specify whether the merger is active at the particular analysis time $t_a$. To resolve this ambiguity, we introduce a key concept in our approach: the `probability of overlap' $p$, which we define as the probability that the merger is actively occurring at, i.e.~overlapping with, the analysis time $t_a$:

\begin{equation} \label{eq:p}
p = 
\max \left\{ \min \left.
\begin{cases}

\frac{T/2-(t_a-t_{m, e})}{t_{m, e}-t_{m, s}}, & {t_{m, e} \le t_a} \\
\frac{T/2-(t_{m, s}-t_a)}{t_{m, e}-t_{m, s}}, & {t_{m, e} > t_a}

\end{cases} , 1
\right\} , 0 \right\}
\end{equation}

Eq.~\ref{eq:p} implements simple principles. Any $t_a$ that is less than $T/2$ away from both $t_{m, s}$ and $t_{m, e}$ has a 100\% probability to be covered by the observability window of the merger, because in this case any point that is between $t_{m, s}$ and $t_{m, e}$, namely any point that is potentially the actual occurrence time of the merger, is also less than $T/2$ away from $t_a$. In other words, $t_a$ is necessarily less than $T/2$ away from the middle point of the observability window, and therefore necessarily encompassed by it. On the other extreme, any $t_a$ that is more than $T/2$ further away from both $t_{m, s}$ and $t_{m, e}$ has a zero probability to be encompassed by the observability time window, as it is outside the bounding range discussed above. In between these cases, where $t_a$ is closer than $T/2$ to one of $t_{m, s}$ or $t_{m, e}$ but not the other, we should consider the edge of the observability window that is closer to $t_a$: the probability of overlap, namely that $t_a$ is contained within the observability window, is simply the fraction of all possible positions of that edge in which $t_a$ is between that edge and the closer of $t_{m, s}$ or $t_{m, e}$, as demonstrated in Figure \ref{fig:overlap}. In essence, we take this probabilistic approach because we are ignorant about the exact time point during the simulation at which the descendant actually appeared (since the galaxy finder only samples certain snapshots in time), and so we effectively assume a uniform probability density function (PDF) for that appearance between $t_{m, s}$ and $t_{m, e}$.

\subsubsection{Traversing the merger tree}\label{subsubsec:M_mrgrID_traversal}

From Section \ref{subsubsec:M_mrgrID_time}, we see that prospective mergers can occur (namely have their direct descendants) at times other than the analysis time. Therefore, when analyzing a galaxy $G_a$ (selected at the analysis time $t_a$) we must search for overlapping mergers both before and after $t_a$.

For previous mergers, we begin by finding the first progenitor of $G_a$ and then iterate through all its other direct progenitors, termed `next progenitors'. For each of those next progenitors, we calculate the stellar mass ratio $\mu$ with respect to the first progenitor (for details, see Section \ref{subsubsec:M_mrgrID_M}), and confirm the pair as a prospective merger if both $\mu_{\textup{min}} \le \mu \le 1/\mu_{\textup{min}}$ and it has a non-zero probability of overlap $p$. We then move further up the first progenitor branch, continuing iteratively: the previous first progenitor becomes the new descendant, we find the new descendant's first progenitor and next progenitors, and record any prospective mergers, continuing until the earliest time on the first progenitor branch. However, any galaxy resulting from a merger is not just the product of its first progenitor but also of any `next progenitors' that formed mergers with the first progenitor. Therefore, for previous mergers, we check for prospective mergers not only along the original first progenitor branch, but also along the first progenitor branches of all the other progenitors that were already identified as contributing to prospective mergers, whether they are direct progenitors of $G_a$ or further back in time. In doing so, each merger's mass ratio is defined in the usual way, namely based on its own direct progenitors. This means, for example, that with a mass ratio threshold of $\mu_{\textup{min}}=1/4$, an object with progenitors that have a mass ratio of $\mu=1/3$, where the less massive progenitor is itself the product of a merger with a mass ratio of $\mu=1/3$, will be considered a multiple merger system even though the mass ratio between the most massive and least massive objects in the trio is $1/9$.

We use a similar procedure for subsequent mergers, namely those at $t>t_a$, by following the tree along the descendant branch. When the galaxy is the first progenitor of its own descendant, we look for prospective mergers by comparing the followed galaxy to all other progenitors of its descendant. For subsequent mergers, however, the followed galaxy might not be the first progenitor of a set of progenitors (as it might be merging with a larger galaxy). If the followed galaxy is not the first progenitor in a progenitor set, it must merge with the first progenitor; hence in these cases we compare the followed galaxy only to the first progenitor (rather than to other progenitors in the set). We continue this process even if the descendant line of $G_a$ is subsumed into a larger galaxy's tree such that it is no longer the first progenitor of its descendant, unless it is subsumed into a galaxy much more massive than itself (by a factor $> 1/\mu_{\textup{min}}$); otherwise, the process continues until the root of the tree (at $z$ = 0) is reached.

\subsubsection{Virtual progenitors and descendants} \label{subsubsec:M_mrgrID_skip}

Finally, in an effort to further improve the accuracy of our measurements, we attempt to compensate for the subhalo skipping issue noted in RG15, whereby in the SubLink trees the descendant is sometimes allowed to be two snapshots after of the progenitor, which is done in order to preserve the continuity of the tree even if the structure finder 'skips' a single snapshot by failing to identify the galaxy in it. To address those cases, we begin by considering that galaxies generally increase in mass over time due to accretion, but some of that mass is not reflected in the merger trees. For any merger, then, the sum of all $n$ progenitor masses (at their max past masses; see Section \ref{subsubsec:M_mrgrID_M}) $\sum_1^nM_{P_i}$ is expected to be close to, but not exceed, the mass of the descendant $M_D$.

\begin{figure}[ht!]
\centering
\includegraphics[scale=0.8]{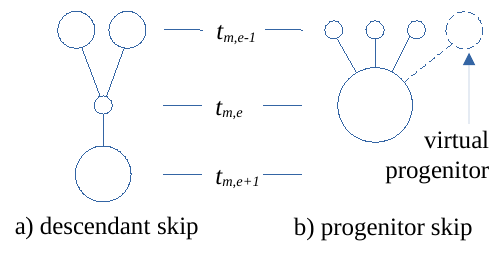}
\caption{Illustration of descendant and progenitor skips. On the left is a descendant skip, where the mass of the descendant (at time $t_{m,e}$) is less than the total mass of its progenitors. Since galaxy masses typically increase over time, this is likely a skip, so we set the mass of the descendant to be equal to that of its progenitors at time $t_{m,e-1}$. On the right, the mass of the progenitors at time $t_{m,e-1}$ is much less than that of its descendant at time $t_{m,e}$. This is a case of a likely progenitor skip, which we remedy by creating at time $t_{m,e-1}$ a `virtual progenitor' galaxy with mass equal to the mass of the descendant that is in excess of the sum of the masses of its progenitors. Since there is some ambiguity with regards to when such a remedy is appropriate (since some mass growth can truly occur via unresolved objects), we generate and compare results both with and without these virtual progenitors.
\label{fig:vp}}
\end{figure}

Therefore, if $\sum_1^nM_{P_i} > M_D$, we assume that the descendant was skipped; to address this, we manually set $M_D = \sum_1^nM_{P_i}$ for the purpose of recording mergers by their descendant mass (see the left side of Figure \ref{fig:vp}). If, on the other hand, the sum of the progenitor masses is much smaller than the descendant mass, $\sum_1^nM_{P_i} / M_D < c \ll 1$ (specifically, we set $c = 0.01$), we assume that the first progenitor was skipped and create a virtual (first) progenitor, with mass $M_{\textup{FP}} = M_D - \sum_1^nM_{P_i}$ (see the right side of Figure \ref{fig:vp}). Since this case is somewhat ambiguous, we generate results both with and without the generation of virtual progenitors.

\subsubsection{Merger mass ratio} \label{subsubsec:M_mrgrID_M}
\begin{figure}[ht!]
\centering
\includegraphics[scale=0.8]{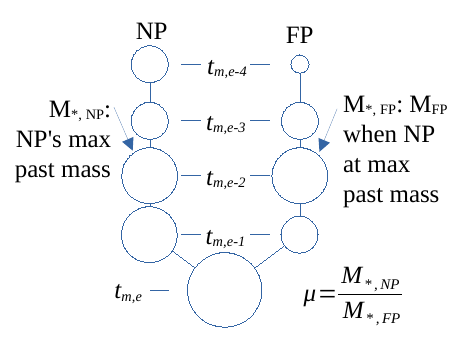}
\caption{An illustration of the maximum past mass algorithm for determining the mass ratio of a merger, following RG15. The smaller progenitor (`next progenitor', here NP) is followed backwards along its first progenitor branch and the time of its maximum past mass is determined (depicted here at $t_{m,e-2}$ with the largest circle along the left branch). Its mass at that time is then compared to the mass of the primary galaxy (`first progenitor', here FP) at that same time, and their ratio is recorded as the merger's mass ratio.
\label{fig:tree_mpm}}
\end{figure}

To assign a (stellar) mass ratio to each prospective merger, we follow the maximum past mass algorithm of RG15, illustrated in Figure \ref{fig:tree_mpm}. In particular, for each analyzed galaxy $G_a$, we follow the tree backward in time to obtain its first progenitor as well as all of its other progenitors (the `next progenitors'). We then follow each potentially merging next progenitor backward in time along its own first progenitor branch until its first appearance in the tree. We then compare the next progenitor's maximum past mass over its previous history with the mass of the first progenitor at the next progenitor's time of maximum past mass, and their ratio is recorded as the merger mass ratio $\mu=M_{\ast, \textup{NP}}/M_{\ast, \textup{FP}}$. For any predefined mass ratio threshold $\mu_{\textup{min}}$ (see Section \ref{subsec:M_param}), the prospective merger is selected for analysis under the condition that $\mu_{\textup{min}} \le \mu \le 1/\mu_{\textup{min}}$.

\subsection{The Merger Fraction} \label{subsec:M_f}

Equipped with the probability of overlap between any given merger event and any given object in the merger tree, we proceed in Section \ref{subsubsec:M_f_measured} to defining and measuring the `merger fraction' $f$, which is shorthand for the fraction of objects in a given galaxy sample at a given time that is actively undergoing a merger (or more than one merger) with some given properties. In Section \ref{subsubsec:M_f_RG15} we review for reference a more traditional way of calculating the merger fraction, which does not allow quantifying multiple mergers.

\subsubsection{Our method: The probability-based merger fraction} \label{subsubsec:M_f_measured}

As we turn to the calculation of the merger fraction, we distinguish between the total merger fraction $f_t$, namely the fraction of galaxies undergoing at least one merger at the analysis time, and the multiple merger fraction $f_m$, namely the fraction of galaxies undergoing at least two mergers at the same time. In each case, $f = N_M / N_G$, where $N_G$ is the number of galaxies in the selected sample, and $N_M$ is the number of those that are actively undergoing at least $M$ mergers. Hence, $f_t=N_1/N_G$ and $f_m=N_2/N_G$.

Our task at hand can then be recast as using the set of merger overlap probabilities $p_i$ associated with each galaxy $G_a$ to find the expectation value of its contribution to $N_M$ for various values of $M$. This requires that all $n$ mergers with a non-zero probability of overlap with $G_a$ be analyzed together. To do this, we create two tables for each $G_a$ (see Table \ref{table:probs} for an illustration) with $n$ columns and $2^n$ rows each, which represent all possible Boolean permutations of $n$ elements. The Boolean values in the `overlap flag table' represent the cases of overlap (1) or non-overlap (0), and the corresponding values in the `probability table' are set to $p_i$ and $1-p_i$, respectively, where $p_i$ is the probability of overlap for merger $i$, given by Eq.~\ref{eq:p}. The example shown in Table \ref{table:probs} is for the case of a galaxy with three mergers $m_1$, $m_2$, and $m_3$, with overlap probabilities $p_1$, $p_2$, and $p_3$.

\begin{deluxetable}{cccccccc}
\tablecaption{Probability table illustration \label{table:probs}}
\tablehead{
\multicolumn{3}{c}{Overlap flag} & \multicolumn{2}{c}{} & \multicolumn{3}{c}{Probability}
}
\startdata
\phantom{1}$m_1$\phantom{1} & \phantom{1}$m_2$\phantom{1} & \phantom{1}$m_3$\phantom{1} & \phantom{} & \phantom{} & \phantom{00}$m_1$\phantom{00} & \phantom{00}$m_2$\phantom{00} & \phantom{00}$m_3$\phantom{00}\\
\hline
0 & 0 & 0 & \phantom{.} & \phantom{.} & $1-p_1$ & $1-p_2$ & $1-p_3$\\
0 & 0 & 1 & \phantom{.} & \phantom{.} & $1-p_1$ & $1-p_2$ & $p_3$\\
0 & 1 & 0 & \phantom{.} & \phantom{.} & $1-p_1$ & $p_2$ & $1-p_3$\\
0 & 1 & 1 & \phantom{.} & \phantom{.} & $1-p_1$ & $p_2$ & $p_3$\\
1 & 0 & 0 & \phantom{.} & \phantom{.} & $p_1$ & $1-p_2$ & $1-p_3$\\
1 & 0 & 1 & \phantom{.} & \phantom{.} & $p_1$ & $1-p_2$ & $p_3$\\
1 & 1 & 0 & \phantom{.} & \phantom{.} & $p_1$ & $p_2$ & $1-p_3$\\
1 & 1 & 1 & \phantom{.} & \phantom{.} & $p_1$ & $p_2$ & $p_3$\\
\enddata
\tablecomments{The probability table used to determine how many mergers each galaxy is simultaneously undergoing, for the case of $n=3$ potential mergers with a non-zero probability of overlap, namely of being `in progress' (observable) during the analysis time $t_a$. The overlap flags $b_i^j$ on the left side of the table indicate for each potential merger $i\in(1,2,3)$ whether $t_a$ is contained within its duration window, so that all scenarios $j\in(1,..,2^n)$ are covered (with $2^n$ rows). The right side of the table lists the corresponding probabilities of all flag values, where $p_i$ is the probability of merger $i$ overlapping with the galaxy in question at $t_a$. The product $p^j$ of all the probabilities in each row $j$ (given by Equation \ref{eq:p_joint}) provides the probability of the scenario depicted by the combination of the values of the overlap flags in that row, which directly determines (as $\sum_i b_i^j$) the number of mergers whose duration windows overlap with the analyzed galaxy at $t_a$. Note that the sum of the probabilities of all rows equals unity by construction.}
\end{deluxetable}

The product of all probabilities in row $j$ is the joint probability of that Boolean permutation that represents a particular combination of which mergers overlap with $G_a$ and which do not:
\begin{equation} \label{eq:p_joint}
p^j = \prod_{i=1}^{n} p_i^{b_i^j} (1-p_i)^{1-b_i^j},
\end{equation}
where $b_i^j$ is the Boolean value associated with merger $i$ in row $j$. Correspondingly, the number of overlapping mergers represented by row $j$ is equal to $\sum_i b_i^j$. Hence, the total probability for the set of mergers to satisfy a certain condition (e.g.~at least one overlap for $f_t$ and at least two for $f_m$), which is taken as the expectation value of the contribution of $G_a$ to that condition, is the sum of the joint probabilities of all rows corresponding to that condition. By summing up those expectation values from all galaxies $G_a$ in a certain sample, we obtain $N_M$ for that sample:
\begin{equation} \label{eq:N_M}
N_M = \sum_{G_a}\sum_{j|\sum_i b_i^j>=M} p^j.
\end{equation}
This is then straightforwardly converted into a merger fraction by dividing by $N_G$.

For comparison with the total and multiple merger fractions, we also generate merger fractions for particular numbers of overlapping mergers. As noted, while a galaxy contributes to the total merger fraction $f_t$ if it is undergoing \emph{at least} one merger at the analysis time, and to the multiple merger fraction if \emph{at least} two, we also define the `binary` merger fraction, to which only galaxies undergoing \emph{exactly} one merger are contributing, a `tertiary` merger fraction to which only galaxies undergoing \emph{exactly} two merger are contributing, and so on. To do so, we select only the appropriate rows from the probability table for each condition, sum up their probabilities $p^j$, sum up for all galaxies $G_a$, and divide by $N_G$.

To calculate error bounds for each merger fraction measurement, we make the approximation that all galaxies contribute equally to the sum $N_M$ and therefore use the Wilson score intervals for Binomial proportion with continuity correction \citep{https://doi.org/10.1002/(SICI)1097-0258(19980430)17:8<857::AID-SIM777>3.0.CO;2-E}, namely the lower and upper error bounds on $f$ are, respectively,
\begin{equation} \label{eq:Wilson_error}
\begin{split}
& f^- = \max \{ 0, \frac{2N_G \hat{p} + z^2 - [z \sqrt{a + b} + 1]}{2(N_G + z^2)} \} \\
& f^+ = \min \{ 1, \frac{2N_G \hat{p} + z^2 + [z \sqrt{a - b} + 1]}{2(N_G + z^2)} \} \\
\end{split}
\end{equation}
where $a = z^2 - 1/N_G + 4N_G \hat{p} (1 - \hat{p})$, $b = (4 \hat{p} - 2)$, $\hat{p} = f = N_M/N_G$, and $z = 1$ for errors representing one standard deviation.

\subsubsection{The RG15-based merger fraction} \label{subsubsec:M_f_RG15}

RG15 measured the galaxy merger rate in the Illustris simulation following previous work performing analogous measurements on dark matter halos in the Millennium simulation \citep{2008MNRAS.386..577F,2009ApJ...701.2002G}. Its primary target is the merger rate $R$, defined as the number of mergers per galaxy per unit time. However, this quantity is not suitable for considering multiple mergers for two main reasons. First, a multiple merger rate is not well defined in RG15's merger rate formalism because mergers are implicitly assumed to occur instantaneously, while a multiple merger requires at least three galaxies to be merging at the same time, implying that a nonzero merger duration is required. Second, the RG15 method does not distinguish between individual galaxies but rather relies on the overall numbers of mergers and galaxies (for a certain choice of parameters like redshift and mass ratio). It is precisely to overcome these limitations that we have developed the method introduced in Section \ref{subsubsec:M_f_measured} to calculate the merger fraction $f$, where $f$ is the fraction of galaxies undergoing a certain number of concurrent mergers at a given time, divided by the total number of galaxies that exist at that time.

Nevertheless, it is useful to compare our results with those of RG15, and we do so as follows. We first reproduce RG15's method, which calculates the merger rate by determining the total number of mergers occurring within a given time window $\Delta t$ and mass bin, and dividing it by $\Delta t$ as well as by the total number of galaxies:
\begin{equation} \label{eq:R_RG15c}
R_{\textup{RG15}} = \frac{N_{M,\textup{RG15}}}{N_G (t_a - t_{a-1})}
\end{equation}
where $R_{\textup{RG15}}$ is the RG15-based merger rate, $t_a$ is the analysis time (see Section \ref{subsubsec:M_mrgrID_time}), $t_{a-1}$ is the time associated with the snapshot immediately prior to that of $t_a$, $N_G$ is the number of galaxies in the analyzed sample at $t_a$, and $N_{M,\textup{RG15}}$ is the number of mergers whose direct descendant is one of the galaxies in that sample at $t_a$.

To make a direct comparison to our derivation of the merger fraction in Section \ref{subsubsec:M_f_measured}, we convert the RG15 merger rate to an approximate merger fraction by multiplying it by the same merger duration $T$ that we employ in our own method,
\begin{equation} \label{eq:fRG15c}
f_{t, \textup{RG15}} \equiv R_{\textup{RG15}} T = \frac{N_{M,\textup{RG15}} T}{N_G (t_a - t_{a-1})},
\end{equation}
where $f_{t, \textup{RG15}}$ is to be compared to our $f_t$ from Section \ref{subsubsec:M_f_measured}.

To derive error bounds for $f_{t, \textup{RG15}}$, we use a variation on Eq.~\ref{eq:Wilson_error}. In this case, we cannot use the measured fraction for $\hat{p}$, as per Eq.~\ref{eq:fRG15c} it includes $N_{M,\textup{RG15}}$, $N_G$, $T$, and $t_a - t_{a-1}$, but we want only the error associated with $N_{M,\textup{RG15}}$ and $N_G$. To that end, we set $\hat{p}$ to $N_{M,\textup{RG15}}/N_G$, and multiply the resulting error bounds by the same $T / (t_a - t_{a-1})$ as used in Eq.~\ref{eq:fRG15c}.

\subsection{Measuring the times to the closest merger and the second-closest merger}
\label{subsec:M_dt}
While measuring the galaxy merger fraction as described in Section \ref{subsec:M_f} is the primary goal of this work, we also wish to analyze the distribution of times between successive galaxy mergers, as the two quantities are closely related. In fact, we argue below that the former can be derived from the latter, given an assumed merger duration. Our first step in doing so is to determine for each object $G_a$ in the merger tree the time difference between its selection time $t_a$ and the occurrence time of its closest (as well as second-closest) merger -- either along its progenitor or descendant branches. Hereafter, we will use $\Delta t_1$ for the time to the closest merger, $\Delta t_2$ for the time to the second-closest merger, and $\Delta t$ as a more general notation that can incorporate either, as appropriate.

That the actual occurrence time of a merger carries an uncertainty, as it could be anywhere in the range $[t_{m, s}, t_{m, e}]$ as discussed in Section \ref{subsubsec:M_mrgrID_time}, propagates to an uncertainty in the determination of a galaxy's $\Delta t$. We address this in a way that is consistent with our measurement of the merger fraction, namely by using the permutation tables introduced in Section \ref{subsubsec:M_f_measured}. For each galaxy with at least one merger with a non-zero probability of overlap, we extend the permutation table with determinations for when in the window $[t_{m, s}, t_{m, e}]$ we consider each merger to have actually occurred. Specifically, for entries in the table that correspond to an overlap of a merger with $t_a$, namely those where $b_i^j=1$, we make the assumption that the merger occurs at whichever of $t_{m, s}$ or $t_{m, e}$ that is closer to $t_a$. Conversely, if the table element indicates that there is no overlap between $t_a$ and the merger, then its occurrence is tagged at whichever of $t_{m, e}$ or $t_{m, s}$ that is farther away from $t_a$.

For a specific permutation of Boolean values (i.e.~row in the permutation table), we therefore have a set of calculated time differences between $t_a$ and each merger $i$ in the row. Therefore, $\Delta t_1$ is the smallest value in this set and $\Delta t_2$ is the second smallest. The probabilities of overlap $p_i$ from Section \ref{subsubsec:M_mrgrID_time} are again used to set the probabilities $p^j$ for each row of the table (based on Equation \ref{eq:p_joint}) and therefore the probabilities for the values of $\Delta t_1$ and $\Delta t_2$. From the obtained values of $\Delta t_1$ and $\Delta t_2$ and their associated probabilities, we construct a two-dimensional histogram of $\Delta t_1$ and $\Delta t_2$ over all galaxies $G_a$ in a sample. We also generate one-dimensional histograms of the time to the (second-)closest mergers separated into the cases of previous mergers and subsequent mergers, which allows us to investigate the relative influence of previous and subsequent mergers on $\Delta t$. These histograms can then be normalized to serve as effective (marginalized or joint) PDFs.

Finally, in addition to the intrinsic interest in these distributions, the determination of the time-to-closest-merger is of interest because of its relation to the merger fraction. Under the assumption from Section \ref{subsubsec:M_mrgrID_time} that mergers occur in the middle of their associated merger duration $T$, it is the case that the merger fraction equals the integral of the PDF of $\Delta t$ between $0 \le \Delta t \le T/2$. This is because this integral represents the fraction of the total population that has its closest merger at a distance of $\le T/2$, and therefore for which the closest merger overlaps with $G_a$. This holds both for the case of the total merger fraction, with $f_t$ derived from the distribution of $\Delta t_1$, and for the multiple merger fraction, with $f_m$ derived from the distribution of $\Delta t_2$. Beyond being valuable on its own as an alternative method to measure the galaxy merger fraction, $\Delta t$ distributions also allow us, as discussed next, to gain insight by comparing the actual distributions inferred from the simulations with theoretical toy model distributions and their associated merger fractions.

\subsection{Toy models for the distributions of times to the closest and the second-closest mergers} \label{subsec:M_fits}

To gain some perspective into the simulation results, we develop comparison toy scenarios for the distribution of merger events through time and among galaxies. For each of these scenarios, we analytically derive the distributions of the times to the closest and second-closest mergers, and from those, we derive the associated merger fractions, following Section \ref{subsec:M_dt}.

Each of the two scenarios described in Sections \ref{subsubsec:M_fits_dtc} and \ref{subsubsec:M_fits_P} requires the overall merger rate $R$ as an input parameter, for which we use our calculation that follows the RG15 method, as seen in Section \ref{subsubsec:M_f_RG15}. In addition, they require an assumption about how the occurrence of mergers is distributed within the galaxy population. For this, we introduce two flavors to each of the scenarios in Sections \ref{subsubsec:M_fits_dtc} and \ref{subsubsec:M_fits_P}. On one end of the spectrum, we make the extreme assumption that all galaxies experience the same merger rate on average. The alternative scenario we assume is that galaxies are distributed equally but only among those galaxies that {\it in the simulation} experience any mergers at all, at any point in their evolution history, the fraction of which we denote $F$. In this second type of toy scenario, then, a fraction $1-F$ of galaxies never experience a merger, while the effective merger rate for a fraction $F$ of all galaxies is $R/F$, and it is this enhanced rate that we are taking into consideration.

\subsubsection{Constant time between mergers} \label{subsubsec:M_fits_dtc}

\begin{figure}[ht!]
\centering
\includegraphics[scale=0.8]{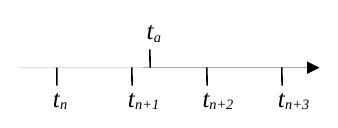}
\caption{Illustration of the toy scenario where mergers occur at constant intervals on a galaxy's timeline. Here $t_{n+1} - t_n = t_{n} - t_{n-1}$ for any $n$ and the merger rate is therefore $R=1/(t_{n+1} - t_n)$. An analysis time $t_a$ is marked at an arbitrary location, representing the independence between the arbitrary times at which simulation snapshots are written and the merger occurrence sequence.
\label{fig:dt_Rc}}
\end{figure}

Our first toy model for the distribution of merger occurrence times assumes a constant fixed time between mergers. We consider an infinite succession of equally spaced mergers $t_1 ... t_n$ (where $t_{n+1} - t_n = R^{-1}$) and a randomly placed analysis time $t_a$, as in Figure \ref{fig:dt_Rc}. 

We first consider the time-to-closest-merger, $\Delta t_1$. When $t_a = t_n$, we have simply $\Delta t_1=|t_a - t_n|=0$. As $t_a$ is placed progressively to the right, $\Delta t_1$ increases linearly until $t_a = (t_{n+1} + t_n) / 2$, when its value reaches a maximum of $(t_{n+1} + t_n) / 2 - t_n = (2R)^{-1}$ (or $(2R/F)^{-1}$, as appropriate), and then decreases linearly back to 0 when $t_a = t_{n+1}$. Beyond $t_a = t_{n+1}$, the cycle repeats infinitely. The PDF of $\Delta t_1$ assuming a constant time between mergers is therefore zero everywhere except when $0 \le \Delta t_1 \le (2R)^{-1}$ (or $0 \le \Delta t_1 \le (2R/F)^{-1}$, as appropriate), and its density there must be uniform. For $\Delta t_2$, a similar consideration gives a PDF of zero except in the range $[(2R)^{-1}, R^{-1}]$ (or $[(2R/F)^{-1}, (R/F)^{-1}]$), where it is uniform.

\subsubsection{Merger occurrence as a Poisson process} \label{subsubsec:M_fits_P}

Next, we model the time distribution of mergers by assuming that they occur independently and at random with a fixed rate, i.e.~that mergers occur as a Poisson process. We can therefore employ the Erlang distributions, such that an exponential distribution (Erlang with shape parameter $k=1$) gives the PDF of the time from the analysis time to that of the closest merger, and an Erlang distribution with shape $k=2$ describes the PDF of the time to the second-closest merger. It is worth noting that since the Erlang distributions provide the time to the \emph{next} event (merger), while we are interested in the time to both next \emph{and previous} events, in both cases we set the rate parameter of the assumed Poisson process to $2R$ (or $2R/F$, as appropriate), leveraging the time symmetry of the Poisson process.

\subsection{Parameter Space} \label{subsec:M_param}

We vary multiple parameters in an attempt to determine their effect on the results. First, as noted above, our work is based on the galaxy merger trees of the Illustris and IllustrisTNG simulations. Specifically, we generate results using the merger trees from four simulations: Illustris-1, Illustris-3, TNG100-1 (hereafter TNG100) and TNG300-1 (hereafter TNG300). Illustris-1 and Illustris-3 are two of the three Illustris full-physics simulations, with Illustris-1 the flagship simulation and Illustris-3 the lowest resolution of the three baryonic simulations. TNG100 and TNG300 are two of the three flagship IllustrisTNG simulations. TNG300 is the largest volume of the three, but with the lowest resolution; TNG100 is meant to balance both resolution and volume and has the same initial conditions as Illustris-1, allowing results between the two to be directly compared. As mentioned in Section \ref{subsec:M_merger_ID}, we employ two versions of SubLink, a dark matter-based version (`SubLink'), and a baryonic-based one (`SubLink\_gal'). In all four simulations, both merger trees are available, and we use both for completeness and comparison.

For the merger duration $T$, we adopt the one given in \citet{2017MNRAS.468..207S} as

\begin{equation}
\label{eq:T}
T_{\textup{ref}} = 2.4 (1+z)^{-2}\Gyr,
\end{equation}
which shows a strong time dependence that reflects the evolution of the dynamical time in an expanding universe. This formula is independent of mass or mass ratio, for simplicity, and does not incorporate scatter that could be driven e.g.~by orbital differences, however since this quantity is of critical importance in this work, we explore variations of $T$ by multiplying it by a constant $T_{\textup{fac}}$ that takes values of 0.5, 1, or 2. In addition, we explore two variants to address an ambiguity regarding whether the input redshift in Equation \ref{eq:T} should be that of the analysis time $t_a$ or the merger time $t_{m, e}$ (setting $T$ based on $t_a$ results in faster, more straightforward analysis, while setting it based on $t_{m, e}$ may be more physically appropriate). We therefore generate results using both methods and compare them. 

Finally, we generate results for ten redshifts: $z$ = 5, 4, 3, 2.5, 2, 1.5, 1, 0.5, 0.2, and 0.1. For each, we select for analysis all galaxies with a stellar mass above $M_{\ast}=10^{9} M_{\odot}$ at the snapshot closest to the analysis redshift, provided that they belong to merger trees that have a final descendant at $z = 0$, and divide them into various galaxy mass bins. We explore three values of the minimum mass ratio $\mu_{\textup{min}}$ (see Section \ref{subsubsec:M_mrgrID_M}): 1/2, 1/4, and 1/10. We also generate results both with and without the virtual progenitors introduced in Section \ref{subsubsec:M_mrgrID_traversal}. Therefore, overall, we compute the various statistics described in this section for a large number of conditions, namely all the combinations of these various parameter space choices.

We also define what we have, somewhat arbitrarily, chosen as the 'fiducial' case for the purposes of presentation in this paper: $z = 2$ merger fractions based on SubLink\_gal trees from TNG100, for merger minimum mass ratios of $\mu_{\textup{min}} = 1/4$, where our virtual progenitors algorithm is turned on and the merger duration is defined based on $T_{\textup{ref}}$ with $T_{\textup{fac}}=1$. Where relevant, namely, where results from only a single mass bin are presented, we choose the fiducial mass bin to be that which includes $M_{\ast}$ = $10^{10} M_{\odot}$.

\section{Results}
\label{sec:results}

\subsection{Merger Fractions}
\label{subsec:R_f}

\subsubsection{Simulation results} \label{subsubsec:R_f_measured}
\begin{figure}[ht!]
\includegraphics[width=0.47\textwidth]{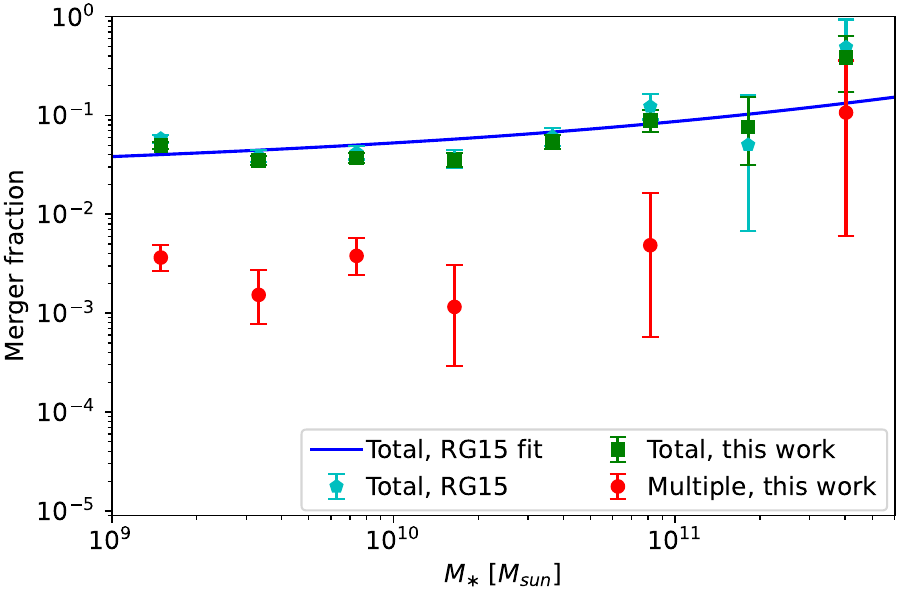}
\caption{Merger fractions versus galaxy stellar mass as measured from the simulations with different methods at the fiducial point in parameter space. The fraction $f_{t, \textup{RG15}}$ derived from TNG100 following RG15 (cyan) is in good agreement with the RG15 fitting function $f_{t, \textup{RG15\ fit}}$ (blue) even though the latter was derived by RG15 for the original Illustris simulation, lending confidence to our implementation of the RG15 method. There is also very good agreement between the total merger fraction $f_t$ derived with our approach (green) and that of the RG15 methods, again providing a good check on our approach. One of the key results of this work is that the multiple merger fraction $f_m$ (red) is consistently 1-1.5 dex below the total merger fraction $f_t$.}
\label{fig:R_fvm_measured}
\end{figure}

Figure \ref{fig:R_fvm_measured} shows the total and multiple merger fractions versus stellar mass for the fiducial case. We begin by discussing the total merger fraction as a useful reference for the multiple merger fraction. Figure \ref{fig:R_fvm_measured} presents two calculations of the total merger fraction: $f_t$ (green), which uses the method introduced in this work, and for comparison $f_{t, \textup{RG15}}$ (cyan), which uses the method described in RG15, along with the fitting function of RG15 $f_{t, \textup{RG15\ fit}}$ (blue curve). Importantly, there is generally good agreement between $f_t$ and $f_{t, \textup{RG15}}$, providing a good check on our method and its implementation.

\begin{figure}[ht!]
\includegraphics[width=0.47\textwidth]{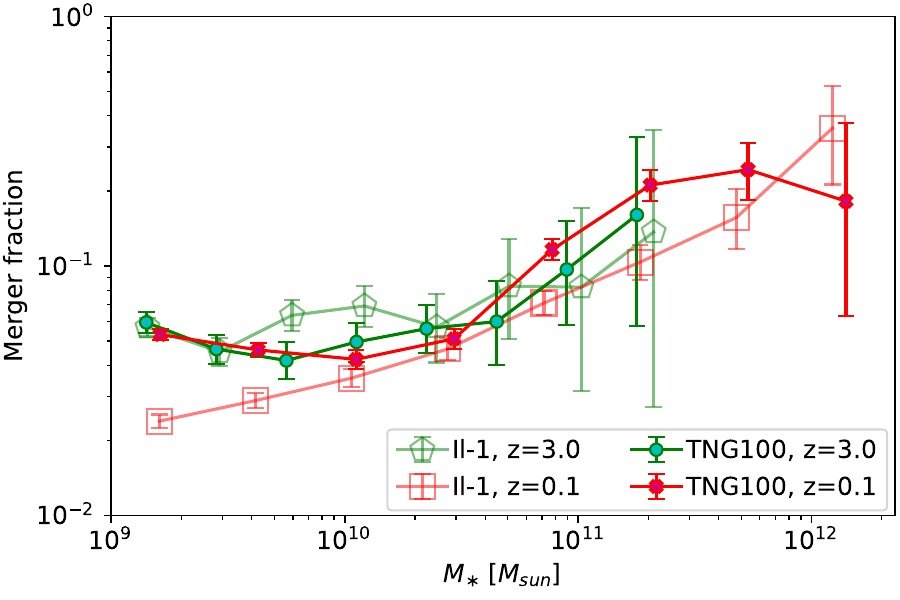}
\caption{A comparison of the merger fractions measured from TNG100-1 (dark colors) and Illustris-1 (light colors), as a function of stellar mass. Notable differences are the greater second derivative of the mass dependence of the TNG merger fractions, and the tendency towards overall higher fractions in TNG, compared to Illustris, towards lower redshifts.}
\label{fig:R_fvm_measured_TNG_Ill1_comparison}
\end{figure}

Figure \ref{fig:R_fvm_measured} also shows that the merger fractions we calculated are reasonably well described by the fitting function $f_{t, \textup{RG15\ fit}}$, which provides a good check on our implementation of the RG15 method. However, the difference between the two is systematic and can be described as $f_{t, \textup{RG15}}$ having a larger second derivative than $f_{t, \textup{RG15\ fit}}$. Much of this difference can be traced to the difference between the simulations, namely that our chosen fiducial case for $f_{t, \textup{RG15}}$ uses the TNG simulation while $f_{t, \textup{RG15\ fit}}$ was derived for the original Illustris simulation. This can be seen explicitly in Figure \ref{fig:R_fvm_measured_TNG_Ill1_comparison}. We also note that the merger fractions in the original Illustris simulation decrease toward lower redshifts faster compared to those in the TNG simulation, with deviations reaching up to $\sim0.5$ dex at low redshift, on average. While a detailed account of this difference is beyond the scope of this paper, it is qualitatively consistent with expectations based on the distinct evolutions of the stellar mass to halo mass ratio in these two simulations. In particular, the original Illustris shows higher ratios than in TNG, more and more so towards low redshifts \citep{Pillepich_2017a}, implying that at a fixed stellar mass, Illustris galaxies reside in less massive halos compared to TNG ones, more so towards lower redshift. Given that lower mass halos have lower merger rates \citep{2009ApJ...701.2002G}, a faster decline with cosmic time of the galaxy merger rate in Illustris compared to TNG is to be expected.

Examining the multiple merger fraction $f_m$ in Figure \ref{fig:R_fvm_measured} (red), we find that it is consistently 1-1.5 dex below the total merger fraction $f_t$. At high mass ($M_{\ast} \geq 10^{11} M_{\odot}$), $f_m$ tends to steepen towards $f_t$. That $f_m$ tracks $f_t$ with a steepening at high masses is also typical in other regions of the parameter space, not only in the fiducial case. As parameter values are varied, the difference ($f_t-f_m$) tends to be approximately proportional to the minimum merger ratio $\mu_{\textup{min}}$ (such that it is greater for more major mergers) and inversely proportional to the merger duration factor $T_{\textup{fac}}$ (such that it is greater for shorter merger durations). More quantitatively, as $\mu_{\textup{min}}$ decreases from 1/2 to 1/10 or $T_{\textup{fac}}$ increases from 0.5 to 2, the average difference between $f_t$ and $f_m$ decreases by 0.5 dex.

\begin{figure}[ht!]
\includegraphics[width=0.47\textwidth]{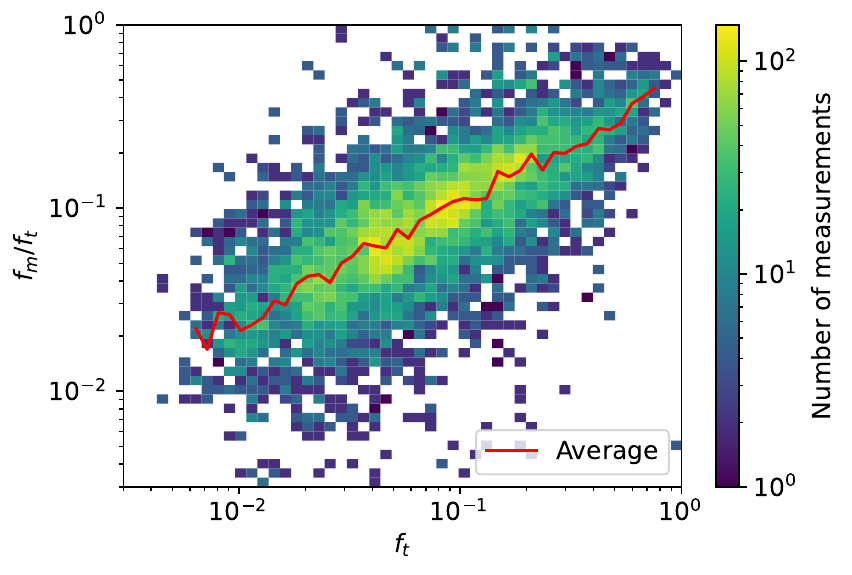}
\caption{A heatmap of the ratio of the multiple to total merger fraction $f_m/f_t$ versus $f_t$, as measured from the simulations across our full parameter space, namely all galaxy mass bins, redshifts, merger minimum mass ratios, merger tree flavors, merger duration assumptions, etc. The red curve represents a running average. Albeit with some scatter, the results are best fit as $f_m/f_t\approx 0.5f_t^{2/3}$.}
\label{fig:ratio_lcF}
\end{figure}

We provide a more general view of our results across the parameter space by presenting in Figure \ref{fig:ratio_lcF} the multiple-to-total merger fraction ratio ($f_m/f_t$) versus the total merger fraction $f_t$ itself, where every point that contributes to the heatmap is one location in our parameter space, namely a combination of mass, redshift, minimum mass ratio, simulation, and analysis nuisance parameters such as the assumed merger duration (see Section \ref{subsec:M_param}). Notwithstanding the considerable spread, it is typically the case, in the areas with the highest density of points, that $f_m/f_t \approx 0.5f_t^{2/3}$, namely that the multiple merger fraction scales as $f_m \propto f_t^{5/3}$, a key result of this work.

\begin{figure}[ht!]
\includegraphics[width=0.47\textwidth]{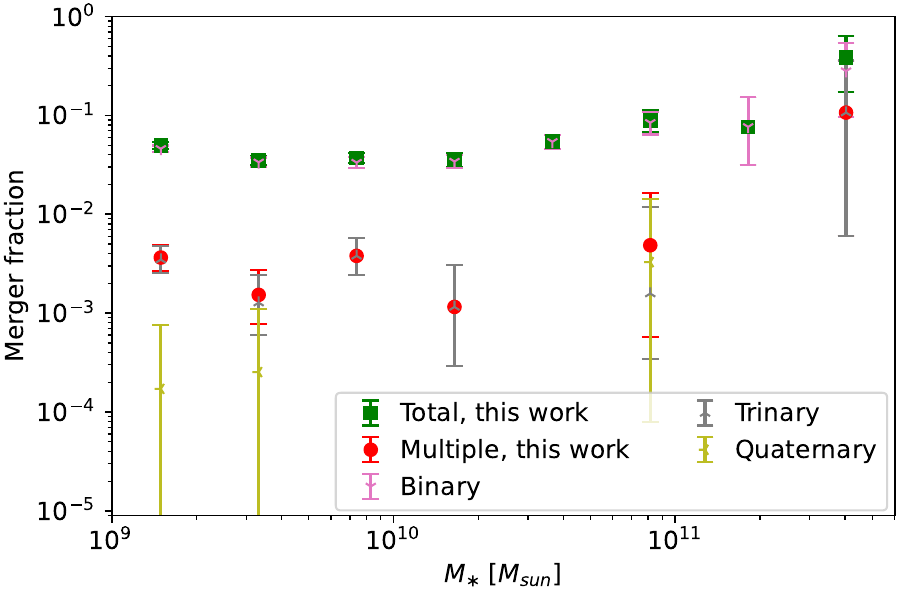}
\caption{Merger fractions versus stellar mass, broken down by the number of mergers the galaxy is undergoing at the analysis time. While the total (green) and multiple (red) merger fractions (repeated from Figure \ref{fig:R_fvm_measured}) include galaxies with \emph{at least} one or two mergers (respectively), the other symbols represent fractions of galaxies undergoing a certain number of concurrent mergers \emph{exactly}: one merger for `binary' (magenta), two for `trinary' (grey), three for `quaternary' (yellow). Across the parameter space generally, and specifically for the fiducial case shown here, the total merger fraction is dominated by the binary fraction and the multiple merger fraction is dominated by the trinary fraction. Also, there is some suggestive vertical compression as the multiplicity increases, e.g.~the logarithmic difference between the binary and trinary fractions is greater than that between the trinary and quaternary} fractions.
\label{fig:R_fvm_ary}
\end{figure}

While the total and multiple merger fractions include galaxies with \emph{at least} one (total) or two (multiple) mergers, we can also quantify the merger fractions of galaxies undergoing \emph{exactly} a certain number of mergers during the analysis time, as noted in Section \ref{subsubsec:M_f_measured}. In Figure \ref{fig:R_fvm_ary}, we present these `binary' merger fraction (exactly one merger; magenta), `trinary' merger fraction (two concurrent mergers; grey), and `quaternary' merger fraction (three concurrent mergers; yellow). For reference, we also show the total (green) and multiple (red) merger fractions, first presented in Figure \ref{fig:R_fvm_measured}. Generally, the total merger fraction $f_t$ is dominated by the binary fraction $f_{{\textup{binary}}}$, and the multiple merger fraction $f_m$ is dominated by the trinary fraction $f_{{\textup{trinary}}}$. The differences between these successive merger fractions generally become smaller as the multiplicity increases, e.g.~$f_{{\textup{trinary}}} - f_{{\textup{quaternary}}}$ tends to be smaller than $f_{{\textup{binary}}} - f_{{\textup{trinary}}}$. These differences also become smaller for longer merger durations and more minor mergers (not shown).

We close this section with a test of potential multiple merger undercounts. With our main analysis, it is possible for a galaxy to be part of a binary merger while the galaxy it is merging with is part of a multiple merger. Here we report on a test of how significant a correction it would be if such galaxies were to be counted as multiple mergers themselves. Specifically, we assess each galaxy that is part of a binary merger to determine if it might be undercounted as part of a multiple merger by checking if its descendant is defined as undergoing a multiple merger according to our usual analysis. For each galaxy, we compute three probabilities: its involvement in a multiple merger, its participation in a binary merger, and its descendant’s involvement in a multiple merger. The product of the latter two probabilities indicates the likelihood that our main analysis undercounts it as a multiple merger. Summing these probabilities of undercounting and comparing to the total multiple merger probabilities, we derive a fractional undercount. We find this fractional undercount is approximately 40\%, implying our main analysis might underestimate the multiple merger fraction by about 30\%. In this sense, our main analysis provides a lower bound for the multiple merger fraction, but since we estimate the undercount factor to be modest, applying it would not qualitatively change any of our results.

Finally, we note that across all merger fraction results, there is little or no appreciable difference between using $T_a$ (the merger duration $T$ determined using the analysis time) or $T_m$ ($T$ determined using the merger time), or between the virtual progenitors algorithm being on or off.

\subsubsection{Toy scenario: merger arrival as a Poisson process} \label{subsubsec:R_f_Poisson}

\begin{figure}[ht!]
\includegraphics[width=0.47\textwidth]{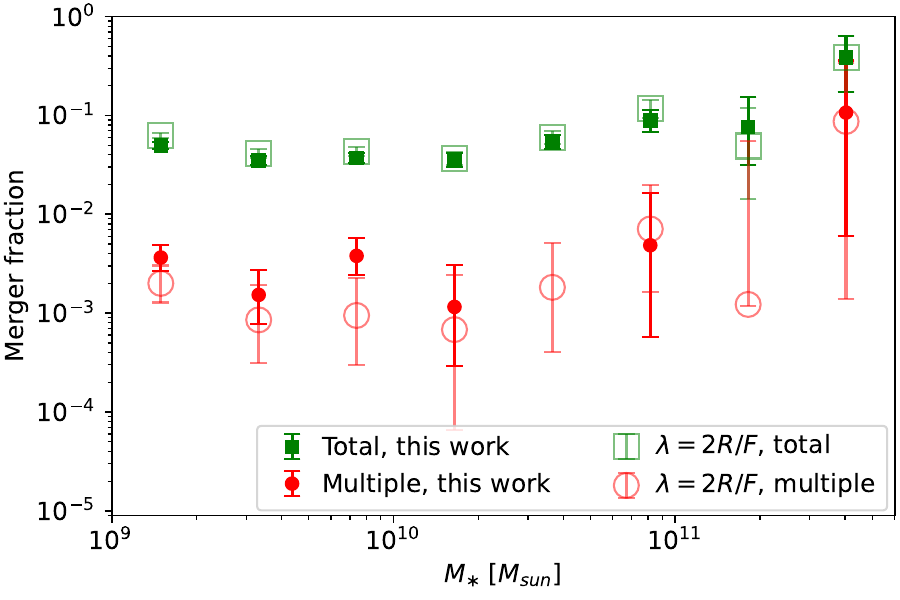}
\caption{Merger fractions versus stellar mass for the toy scenario where mergers occurrence is a Poisson process. The total merger fraction in this scenario (light green) is very close to the actual total merger rate in the simulation (dark green; repeated from Figure \ref{fig:R_fvm_measured}). However, the multiple merger rate in this toy scenario (light red) is $\approx0.5$ dex lower than that measured in the simulation (dark red; repeated from Figure \ref{fig:R_fvm_measured}).}
\label{fig:R_fvm_Poisson}
\end{figure}

As an illuminating comparison to the actual simulation results, in Figure \ref{fig:R_fvm_Poisson} we present results that are obtained by assuming that galaxy mergers arrive as a Poisson process (see Section \ref{subsubsec:M_fits_P}). We convert the distributions of the times $\Delta t_1$ ($\Delta t_2$) to the closest (second-closest) mergers, which are derived using the Poisson process scenario, into total (multiple) merger fractions, respectively, by following Section \ref{subsec:M_dt} and analytically integrating them between $0 \le \Delta t \le T/2$, the result of which is displayed as large empty symbols\footnote{Note that we choose to only show here the case where the mergers are distributed only among the fraction of galaxies that in the simulation have any mergers at all, by considering a rate of $R/F$ among a fraction $F$ of the galaxies, see Section \ref{subsec:M_fits}. This results in a higher multiple merger fraction than the alternative assumption, and therefore serves as an upper limit for the Poisson process scenario.} in Figure \ref{fig:R_fvm_Poisson}. For reference, we also show the measured total (green) and multiple (red) simulation merger fractions, adopted from Figure \ref{fig:R_fvm_measured}. In general, and as seen here for the fiducial case, the measured and Poisson-derived total merger fractions are very close to each other. However, the multiple merger fractions in the Poisson process scenario are typically $\approx0.3-0.5$ dex below those measured from the simulation (with the exception of the highest masses, $M_{\ast} \gtrsim 10^{11} M_{\odot}$).

\begin{figure}[ht!]
\includegraphics[width=0.47\textwidth]{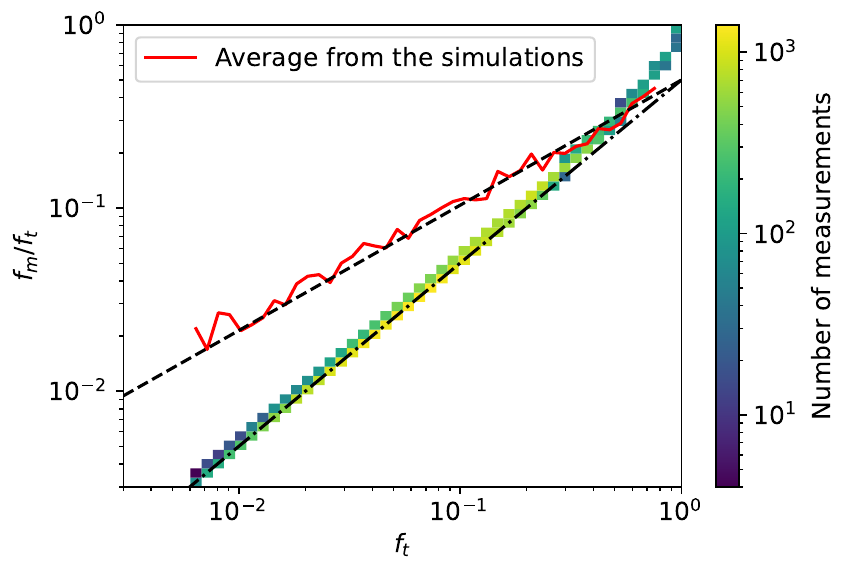}
\caption{A heatmap of the ratio of the multiple to total merger fraction $f_m/f_t$ versus $f_t$, based on the Poisson process toy scenario across the parameter space. At small values, $f_m/f_t\simeq0.5f_t$, namely at a fixed total merger fraction, the multiple merger fraction is smaller than in the actual simulated results shown in Figure \ref{fig:ratio_lcF}, the average of which is repeated here (red). At values approaching unity, the slope increases towards $f_m/f_t$ = $f_t = 1$. The two black lines indicate two power-law relations, one linear that fits the Poisson model and one with an index of $0.68$ that fits the simulation results.}
\label{fig:ratio_lcT}
\end{figure}

Figure \ref{fig:ratio_lcT} compares $f_m/f_t$ to $f_t$ more generally across the parameter space, as in Figure \ref{fig:ratio_lcF}, except here the fractions are those obtained assuming a Poisson process\footnote{We include both results with the input merger rate divided by the fraction of galaxies with mergers $F$ (see Section \ref{subsec:M_fits}), and without doing so. In most cases, the difference between the two is not large since $F\approx1$.}. The results here are essentially free of scatter (compared to those using the measured merger fractions in Figure \ref{fig:ratio_lcF}), and exhibit a clear linear relation (dotted-dashed) at $f_t\ll1$, with $f_m/f_t\simeq0.5f_t$. As $f_t$ increases ($f_t \gtrsim 0.3$), the slope of $f_m/f_t$ versus $f_t$ increases, with $f_m/f_t$ approaching unity as $f_t$ does the same.

We can analytically verify and illuminate the results seen in Figures \ref{fig:R_fvm_Poisson} and \ref{fig:ratio_lcT} for the Poisson process scenario, as follows. The PDF for an exponential distribution with rate $\lambda$, which represents the distribution of the elapsed times to the closest merger ($\Delta t_1$) in this scenario, is 
\begin{equation} \label{eq:pdf1c}
f(t; \lambda)_{\textup{exp}} = \lambda e^{-\lambda t}.
\end{equation}
Recalling that the method we use to determine the total merger fraction of the Poisson process scenario is integrating the $\Delta t_1$ PDF from 0 to $T/2$, by integrating Eq.~\ref{eq:pdf1c} up to $T/2$ and keeping terms only up to the second order we obtain

\begin{equation}
\label{eq:pdf1cfi}
f_{t, {\rm Poisson}} \approx \frac{\lambda T}{2} \left( 1-\frac{\lambda T}{4} \right).
\end{equation}
Per Section \ref{subsubsec:M_fits_P}, the Poisson input rate for closest mergers is

\begin{equation}
\label{eq:l1c}
\lambda = 2R = \frac{2f_{t, \textup{RG15}}}{T},
\end{equation}
therefore, considering that $ f_{t, \textup{RG15}} \ll 1$,

\begin{equation}
\label{eq:pdf1cf}
f_{t, {\rm Poisson}} \approx f_{t, \textup{RG15}} \left( 1-\frac{f_{t, \textup{RG15}}}{2} \right) \approx f_{t, \textup{RG15}},
\end{equation}
which explains the empirical close resemblance noted above between the total merger fraction measured in the simulation and that from the Poisson process scenario.

Similarly, the multiple merger fraction of the Poisson process scenario can be calculated as the integral of the $\Delta t_2$ Erlang PDF of times to the second-closest merger, which equals 

\begin{equation} \label{eq:pdf2c}
f(t; \lambda)_{\textup{Erlang}} = \lambda^2 t e^{-\lambda t},
\end{equation}
the integral of which up to $T/2$ can be approximated (to second order) as

\begin{equation} \label{eq:pdf2cti}
f_{m, {\rm Poisson}} \approx \frac{(\lambda T)^2}{8}.
\end{equation}
Using the same substitution and assumption that led from Eq.~\ref{eq:pdf1cfi} to Eq.~\ref{eq:pdf1cf}, we obtain

\begin{equation} \label{eq:pdf2cf}
f_{m, {\rm Poisson}} \approx \frac{f_{t, \textup{RG15}}^2}{2}.
\end{equation}
Combining Equations (\ref{eq:pdf1cf}) and (\ref{eq:pdf2cf}), we therefore find
\begin{equation}
\label{eq:ratiopdf}
\frac{f_{m, {\rm Poisson}}}{f_{t, {\rm Poisson}}} \approx 0.5{f_{t, {\rm Poisson}}},
\end{equation}
which indeed describes the linear curve with a slope of 0.5 that we see in Figure \ref{fig:ratio_lcT} at small $f_t$. The slope then steepens towards the opposite regime where $f_t$ is large (namely approaching unity), since in that case ${f_{m, {\rm Poisson}}}/{f_{t, {\rm Poisson}}}$ must approach unity too, as it represents a merger rate large enough such that every galaxy is close enough in time not only to the occurrence of one merger but of several.

A comparison between Figures \ref{fig:ratio_lcT} and \ref{fig:ratio_lcF} shows that the multiple merger fractions in the simulations are typically larger than that corresponding to the Poisson process scenario in the vast majority of the parameter space, except where the total merger rate approached unity. The difference between the two is approximately a factor of two at $f_t=0.1$ but grows to approximately a factor of five at $f_t=0.01$ due to the shallower power law that the simulation results display.  We will present additional facets of this basic result in the following sections, but first we discuss here two possible origins for this difference. First, it may emerge from the fact that the merger rate in the simulation is not a constant as is assumed in our Poisson toy model but rather is higher at earlier cosmic times; secondly, it may come about if mergers are not independent of each other as described by a Poisson process, namely if the occurrence of mergers is not locally random in time but rather `clustered' in some way.

\begin{figure}[ht!]
\includegraphics[width=0.47\textwidth]{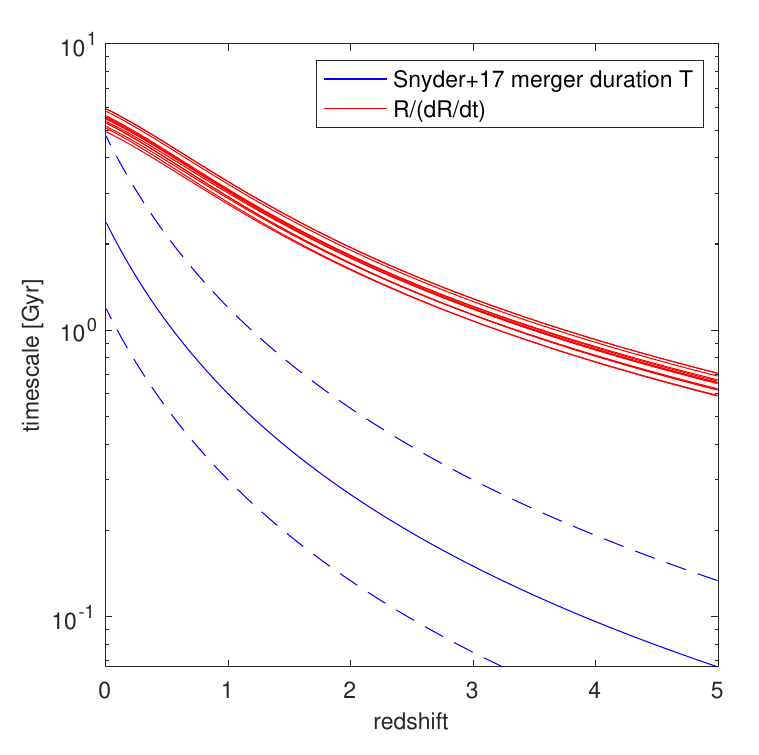}
\caption{A comparison between the merger durations we use in this work (blue), $T_{\textup{ref}}$ (solid), $0.5T_{\textup{ref}}$ and $2T_{\textup{ref}}$ (dashed), and the timescale over which the merger rate evolves (red), which we define as the ratio between the merger rate and its time derivative. The latter is shown for 16 distinct combinations of mass and mass ratio and calculated based on the RG15 formula, demonstrating that it is primarily a function of redshift and not of other parameters. Evidently, throughout our parameter space, the merger rate evolves on longer timescales than the merger duration, making the Poisson process assumption a good one locally at any given time.}
\label{fig:merger_rate_timescales}
\end{figure}

To distinguish between these two possibilities, we hereby present an argument that the former possibility plays at most a minor role, leaving the non-independent occurrence of mergers as the leading explanation. We argue that while a Poisson process by definition has a constant rate, it also serves as a good approximation to a process that has an evolving rate, such as the merger rate in $\Lambda$CDM, as long as the timescale over which the rate is changing is long compared to the relevant other timescales of the problem. In our case, that relevant timescale is the merger duration $T$, since the merger fractions are obtained by integrating the exponential or Erlang distributions up to that value. In Figure \ref{fig:merger_rate_timescales}, we show a comparison between the merger durations we use in this work (blue) and the timescale over which the merger rate evolves (red), based on the RG15 formula. Evidently, throughout our parameter space, the merger rate evolves on longer timescales than the merger duration, making the Poisson process assumption a good one locally at any given time.

We have further verified, but do not show explicitly, that the simulation results themselves as shown in Figure \ref{fig:ratio_lcF} vary only modestly when filtered by the value of $T_{\textup{fac}}$. If the analysis using shorter merger durations ($T_{\textup{fac}}=0.5$) was significantly more similar to a true (constant rate) Poisson process than the analysis using longer merger durations ($T_{\textup{fac}}=2$), over which the merger rate changes more significantly, we would expect to see the results for the former looking more like Figure \ref{fig:ratio_lcT}. However, we do not find that this is the case, implying that the Poisson process toy model deviates from the simulation results primarily because mergers in the simulation are not independent in time as they would be in a Poisson process.

\subsection{The Elapsed Time to the Closest Mergers} \label{subsec:R_dt}

Next, we present the results of our investigation into the distributions of elapsed time between mergers, which are calculated following Section \ref{subsec:M_dt}, in order to support the interpretation of the results from Section \ref{subsec:R_f}. We present our results as a series of histograms of the elapsed time (relative to arbitrarily selected analysis times $t_a$) to galaxies' closest (second-closest) merger, $\Delta t_1$ ($\Delta t_2$). While our main interest here is the relation between $\Delta t_1$ ($\Delta t_2$) and the total (multiple) merger fraction, for completeness we also present separate histograms for past and future mergers, namely for the time since the closest (as well as second-closest) previous merger and for the time until the closest (as well as second-closest) subsequent merger, even though these distinct quantities do not directly relate to the merger fractions by themselves.

To determine the size of the $\Delta t$ bins in these histograms, it is beneficial to consider that, as described in Section \ref{subsec:M_dt}, the fractional abundance of galaxies with $0 \le \Delta t \le T/2$ (where $T$ is the merger duration) is exactly the fraction of galaxies that undergo a merger at $t_a$, according to our definitions. Therefore, in order to make a direct connection between these histograms and the merger fraction, it is important that the range $[0, T/2]$ contains a whole number of $\Delta t$ bins. To achieve that, we set the bin width $w_{\Delta t}$ such that $(T/2)/w_{\Delta t}$ is a whole number, while ensuring that $w_{\Delta t}$ itself is as close as possible to a (somewhat arbitrarily chosen) value of 0.2 Gyr. We then create bins for the $>T/2$ range with the same width $w_{\Delta t}$ as chosen for the $[0, T/2]$ range. In Figures \ref{fig:dt1D1cn0} through \ref{fig:dt1D2cn1}, which present these histograms, we show the value of $T/2$ as a vertical dashed line for reference.

We note that when $w_{\Delta t}$ is small (due to a small $T$), there may be $\Delta t$ bins that are necessarily unoccupied, since the time to the closest mergers can only take values that equal the time separation between the discrete snapshots of the simulation. When the merger duration $T$ is small (in particular at high redshift), the possible values of snapshot separations may be more sparse than the dense $w_{\Delta t}$-sized bins. This presents no fundamental issue, but rather merely gives rise to jagged histograms. For simplicity, in this section we only present results from cases where $T_{\textup{fac}} = 2$ and hence $T$ is not as short as in other cases, which circumvents the issue. We also note that these results are generally insensitive to the hyperparameters of our methodology, such as the merger tree type (SubLink versus SubLink\_gal), the time at which the merger duration is calculated, or the generation of virtual progenitors.

\subsubsection{The elapsed time to the closest merger} \label{subsubsec:R_dt_1D1cn0}

\begin{figure}[ht!]
\includegraphics[width=0.47\textwidth]{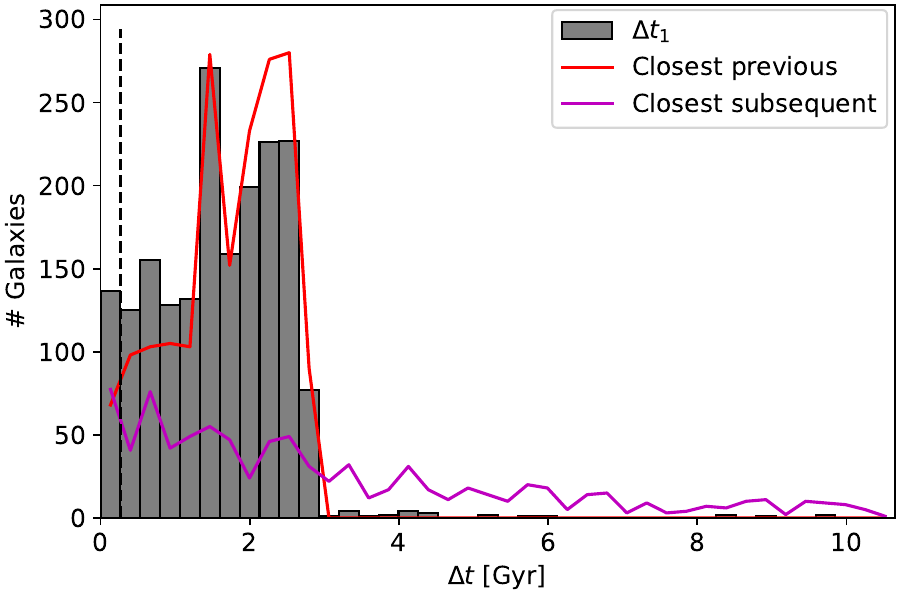}
\caption{Histograms of the time-to-closest-merger $\Delta t_1$, at our fiducial configuration (besides that $T_{\textup{fac}} = 2$ instead of $T_{\textup{fac}} = 1$, for visual clarity reasons; otherwise, the results are essentially the same). The overall $\Delta t_1$ histogram (gray) is presented along with those for the closest previous (red) and closest subsequent (magenta) mergers.  The $\Delta t_1 = T/2$ threshold, which relates this histogram to the merger fraction, is shown by a vertical dashed line.}
\label{fig:dt1D1cn0}
\end{figure}

We begin with Figure \ref{fig:dt1D1cn0}, which presents the histogram of the time to the closest merger for our fiducial parameters. Here, gray bars indicate the number of mergers in the associated $\Delta t_1$ bin. For each bin, we also show in red (magenta) the number of galaxies with the associated time since (until) the closest previous (subsequent) merger. Figure \ref{fig:dt1D1cn0} shows a preferred $\Delta t_1$ value of $\sim2$ Gyr and a sharp cutoff at $\Delta t_1\sim3$ Gyr, corresponding to the age of the universe at our fiducial redshift, and reflecting the finding that the closest merger in most cases is a past merger, due to the decreasing merger rate over cosmic time. We find that across the parameter space, the typical value of $\Delta t_1$ increases with cosmic time, from occupying the smallest bin at $z=5$ to $\Delta t_1\sim10$ Gyr at $z=0.1$. The histograms also tend to flatten and have less distinct peaks toward lower redshift. We finally observe that the relative roles of the previous and subsequent mergers vary with the value $\Delta t_1$, which is discussed further in Section \ref{subsubsec:R_dt_1D2cn0}, as well as with redshift (not shown).

\begin{figure}[ht!]
\includegraphics[width=0.47\textwidth]{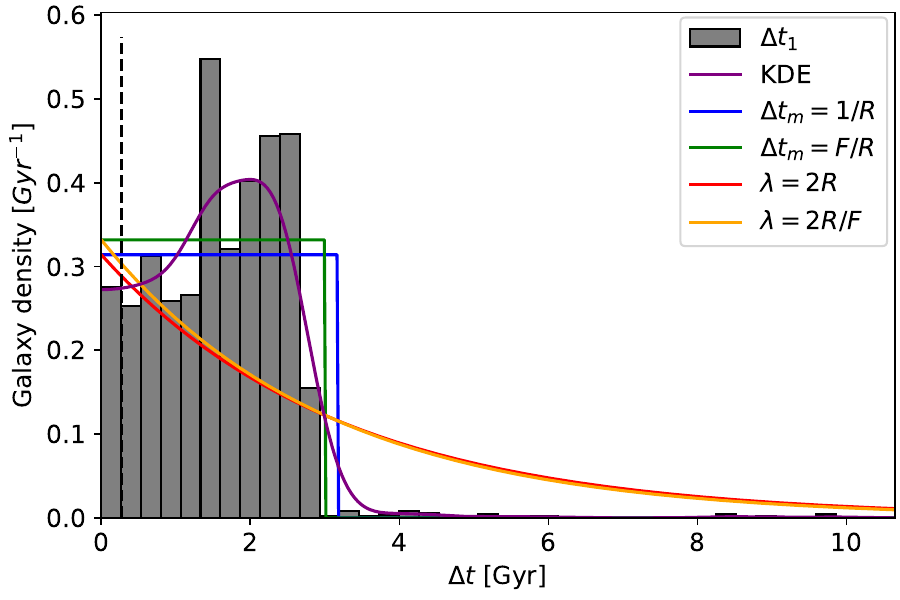}
\caption{PDFs of the time-to-closest-merger $\Delta t_1$ in various scenarios. The simulation results (gray), at our fiducial values (besides that $T_{\textup{fac}} = 2$ instead of $T_{\textup{fac}} = 1$, for visual clarity reasons; otherwise, the results are essentially the same), and their associated Kernel Density Estimate (KDE; purple) are compared against toy models. The model that assumes a constant time between mergers (blue and green, for the cases that do not, or do, respectively, take into account the fact that only a fraction $F$ of galaxies have any mergers with the fiducial mass ratio in their history) provides a reasonable match in this case but not at higher or lower redshifts (not shown). The model that assumes a Poisson process (red and orange, similarly accounting for $F$, respectively) provides a poor fit throughout the parameter space.}
 
\label{fig:dt1D1cn1}
\end{figure}

Next, in Figure \ref{fig:dt1D1cn1} we present the same underlying data in the gray bars, except here normalized to mimic a PDF, instead of a simple histogram, and with the addition of a Kernel Density Estimate (KDE; purple). This allows us to make direct comparisons to the analytical $\Delta t_1$ distributions derived for our toy scenarios described in Section \ref{subsec:M_fits}. The blue and green curves in Figure \ref{fig:dt1D1cn1} correspond to the scenario of constant time between mergers (denoted in the legend as $\Delta t_m$) per Section \ref{subsubsec:M_fits_dtc}, for the cases of the mergers being distributed among the full galaxy population, or among only a subset of galaxies, respectively, as discussed in Section \ref{subsec:M_fits}. The red and orange curves correspond to the Poisson process scenario per Section \ref{subsubsec:M_fits_P}, with its own analogous corresponding flavors.

As Figure \ref{fig:dt1D1cn1} demonstrates, the quantities produced by assuming a constant time between mergers are a somewhat reasonable approximation to the actual simulation results. However, this is not a general result, as we find that at redshifts further away from our fiducial case of $z=2$, the scenario of constant time between mergers shows too peaked distributions (at high $z$) or too wide ones (at low $z$), with respect to the actual simulation results. Furthermore, since the distributions are flat in this toy scenario, they cannot effectively model the peak at the higher $\Delta t$ values (here around $2$ Gyr) that is found in the simulation results. Further, the Poisson process toy scenario evidently fits the measured simulation results even more poorly, as the exponential distribution has a completely different shape, a discrepancy that becomes more significant at lower redshifts. At higher redshifts ($z\gtrsim4$) some parameter configurations do result in a reasonable agreement between the Poisson scenario and the simulation results, as both concentrate towards smaller $\Delta t$ values, but those cases are the exception to the rule.

\subsubsection{The elapsed time to the second-closest merger} \label{subsubsec:R_dt_1D2cn0}

\begin{figure}[ht!]
\includegraphics[width=0.47\textwidth]{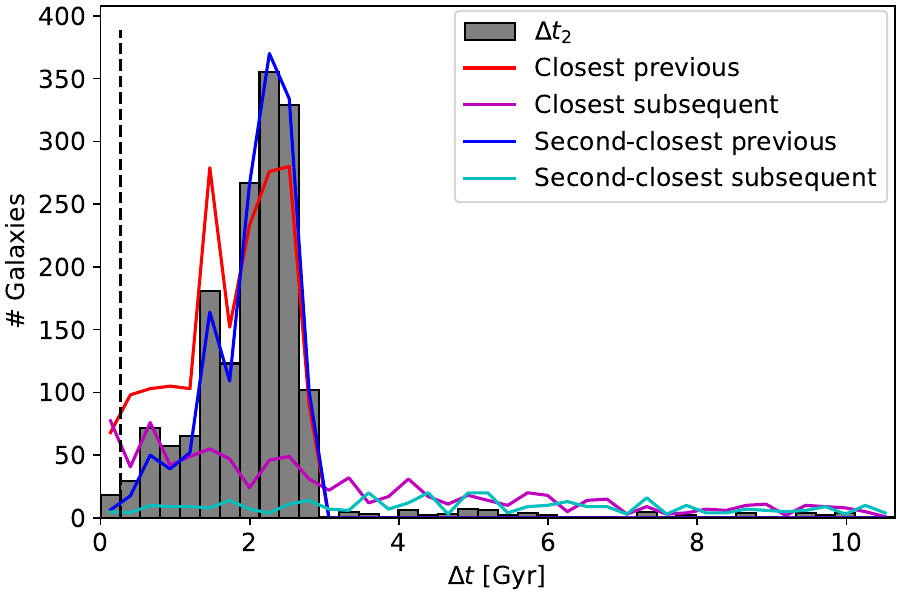}
\caption{Histograms of the times to various close mergers. The histogram of the time elapsed to the overall second-closest merger $\Delta t_2$ (gray) is presented along with distributions of the times elapsed to four other related types of mergers: the closest pre-$t_a$ merger (red), the second-closest pre-$t_a$ merger (blue), the closest post-$t_a$ merger (magenta), and the second-closest post-$t_a$ merger. The histogram area at values of $\Delta t_2 \le T/2$ (left of the vertical dashed line) is of primary interest for the quantification of multiple mergers. In that regime, the second-closest merger is usually the first pre-$t_a$ merger (red). The first post-$t_a$ merger (magenta) usually comes close, but only in the bin containing $\Delta t$ = 0 are they roughly equal.}
\label{fig:dt1D2cn0}
\end{figure}

In Figure \ref{fig:dt1D2cn0} we next present the histogram of the elapsed times to the second-closest merger in gray bars, while the colored curves are repeated from Figure \ref{fig:dt1D1cn0}, with the addition that the blue curve is the time to the second-closest previous merger, and the green is the time to the second-closest subsequent merger. As seen by comparing these two figures that represent the fiducial choice of parameters, as is also the case across the parameter space, the distribution of the time to the second-closest merger spans a similar range to that of the time to the closest merger, but it naturally peaks at larger values. The relative contributions of previous and subsequent mergers to the closest and second-closest merger distributions also vary across the parameter space. However, in the area we are primarily interested in, $0 \le \Delta t_2 \le T/2$, the closest previous and closest subsequent mergers are primarily, and roughly equally, likely to contribute to the closest (and therefore also to the second-closest) mergers.

\begin{figure}[ht!]
\includegraphics[width=0.47\textwidth]{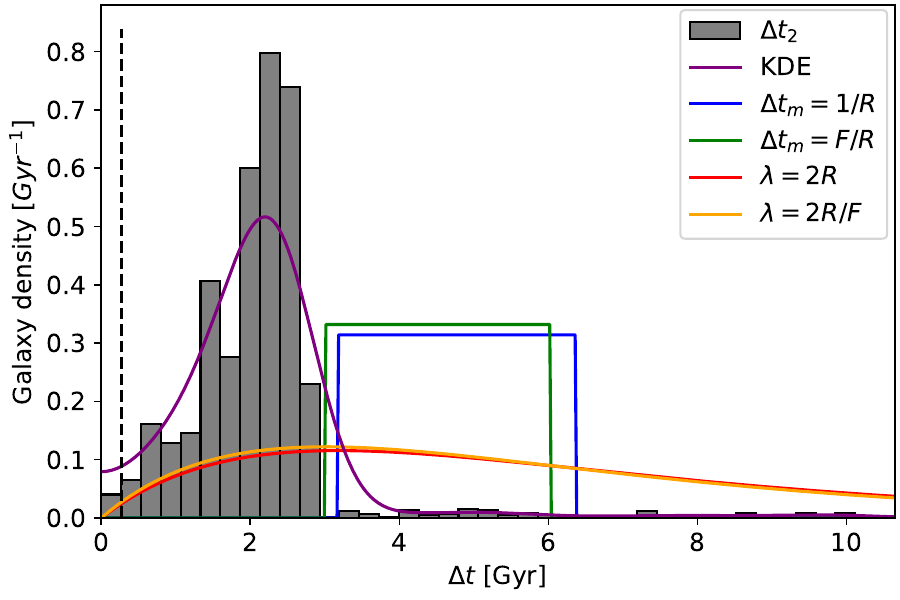}
\caption{PDFs of the time-to-second-closest-merger $\Delta t_2$, analogously to the ones for $\Delta t_1$ shown in Figure \ref{fig:dt1D1cn1}. The simulation results (gray) and their associated KDE (purple) are compared against our two toy models, which provide very poor fits to the data. The toy models, in particular, imply a zero or otherwise low value at $\Delta t_2 \le T/2$, which is directly tied to the multiple merger fraction.}

\label{fig:dt1D2cn1}
\end{figure}

Figure \ref{fig:dt1D2cn1} shows the normalized version of Figure \ref{fig:dt1D2cn0}, including a KDE (purple) and analytical derivations for the toy models. As seen here, and similarly across most of the parameter space, the distributions resulting from assuming a constant time between mergers or a Poisson process do not provide a good match to the measured results. Generally, the toy models result in too extended distributions. In particular, the toy model with a constant time between mergers has zero density at $\Delta t_2 \le T/2$, and therefore corresponds to a zero multiple merger fraction. The results of the Poisson process scenario are generally too low (even if not zero) at $\Delta t_2 \le T/2$, which connects directly to and sheds light on our finding in Section \ref{subsubsec:R_f_Poisson} that the Poisson process scenario results in a lower multiple merger fraction than in the simulations. The Poisson process scenario also has a long tail at high $\Delta t_2$ that does not appear in the simulation results.

Some dependencies on the parameter combination are worth noting. At high $M_{\ast}$, high redshift and small mass ratio $\mu_{\textup{min}}$, namely when the merger fraction tends to be high, the toy model curves are generally narrower and their peak is at smaller $\Delta t_2$ values, as expected. In this regime, the toy model curves are somewhat closer to the simulation results, particularly when $\mu_{\textup{min}}$ is smaller. For the Poisson scenario, this is in agreement with the finding that in this regime there is a greater similarity between the toy model and the multiple merger fractions measured from the simulations (as demonstrated in Figure \ref{fig:ratio_lcT}). For the scenario of constant time between mergers, there exist some input parameters that result in a non-zero density at $\Delta t_2 \le T/2$, especially at small mass ratios $\mu_{\textup{min}}=1/10$ and long merger durations $T_{\textup{fac}}=2$. This implies that this toy model can produce multiple mergers in the high merger rate regime (as opposed to the case shown in Figure \ref{fig:dt1D2cn1}), but this occurs only in the very high-mass end where there are few galaxies and therefore the merger rate is potentially affected by small-number statistics.

\subsubsection{The relation between the times to the two closest mergers}
\label{subsubsec:R_dt_2D}

\begin{figure}[ht!]
\begin{subfigure}{0.98\textwidth}
\includegraphics[width=0.47\textwidth]{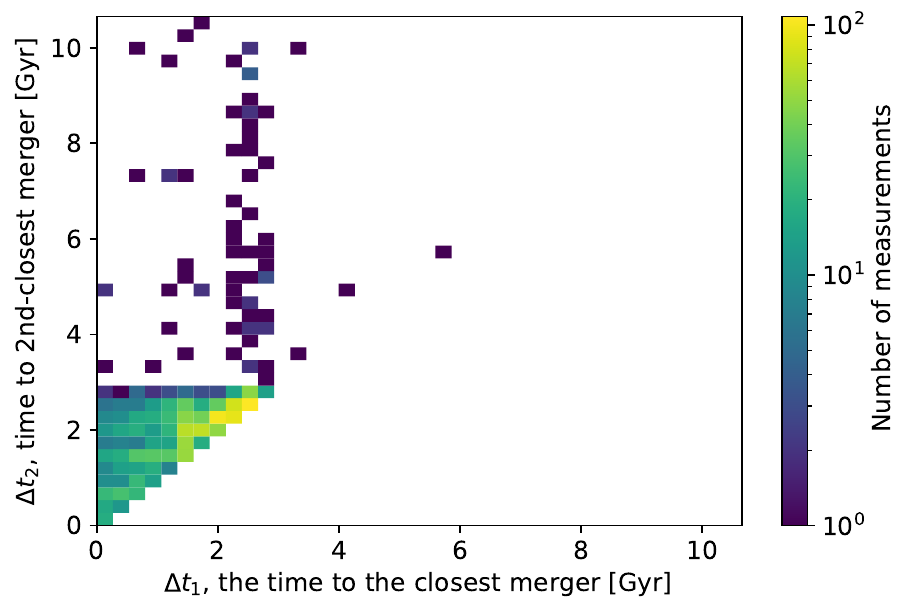}
\label{fig:dt2D_fiducial}
\end{subfigure}
\hfill
\begin{subfigure}{0.98\textwidth}
\includegraphics[width=0.47\textwidth]{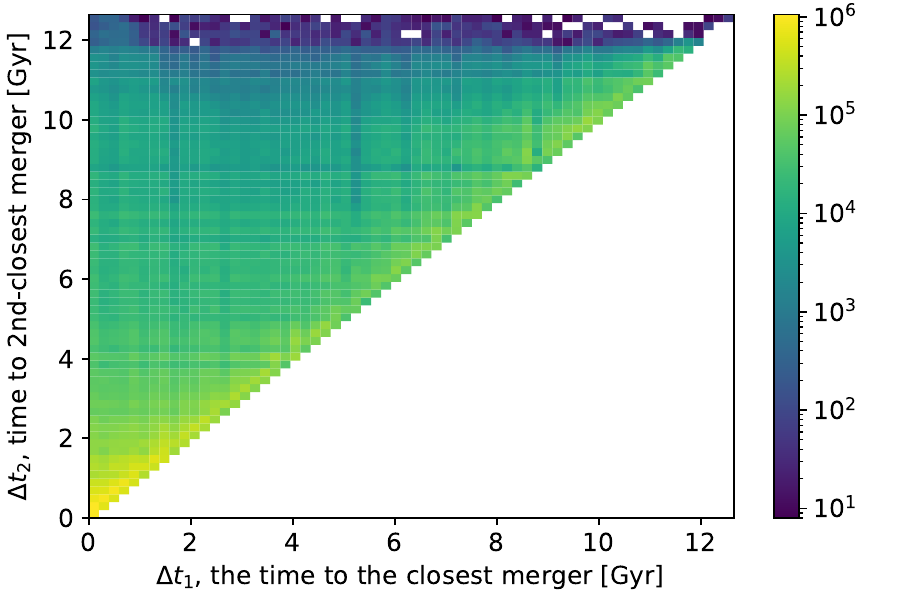}
\label{fig:dt2D_all}
\end{subfigure}
\caption{Heatmaps of the time to the second-closest merger versus the time to the closest merger, at the fiducial parameter values (except for $T_{\textup{fac}} = 2$; top) and combined over the full parameter space (bottom)}. We find that the time until the second-closest merger is typically close to the time to the closest merger, a phenomenon that contributes to the prevalence of multiple mergers.
\label{fig:dt2D}
\end{figure}

Finally, we examine the joint distribution of the time to the closest merger and the time to the second-closest merger. Figure \ref{fig:dt2D} indicates that while (by construction) it is always the case that $\Delta t_2 \ge \Delta t_1$, it is also the case that the distribution of $\Delta t_2$ tends to peak very close to $\Delta t_1$, as indicated by the high-density region just above the $\Delta t_2=\Delta t_1$ line. The majority of the parameter space results in $\Delta t_2 \lesssim \Delta t_1 + 1\Gyr$, and it is only common when $\Delta t_1$ is very small that $\Delta t_2$ is significantly larger. In addition, it is very rare to find values of either $\Delta t_1$ or $\Delta t_2$ that are greater than the age of the universe at the analyzed redshift. These characteristics hold at all redshifts.

\section{Conclusion and Discussion} \label{sec:conclusion}
In this work, we provide the first systematic numerical study of the occurrence of multiple galaxy mergers in cosmological simulations. In particular, we use simulations from the Illustris and IllustrisTNG projects to measure the relationship between the fraction of galaxy systems that are in the process of undergoing a multiple merger, namely a merger comprised of more than two galaxies at the same time, and the total merger fraction, which is dominated by regular, binary mergers. We perform this measurement under various assumptions for what constitutes a merger, encoded primarily in the threshold mass ratio and the assumed merger duration, and for a large range of galaxy masses and redshifts. Our main findings can be summarized as follows.
\begin{itemize}
    \item The typical relation, for some galaxy selection and merger definition, between the total merger fraction $f_t$ and the multiple merger fraction $f_m$, tends to be $f_m \approx 0.5f_t^{5/3}$, albeit with some scatter, and with a potential correction factor that increases it by $\sim40\%$, i.e.~$f_m \approx 0.7f_t^{5/3}$. Since the total merger fraction of any galaxy sample selected by mass and redshift tends to be small (typically $\sim0.1-0.01$), it follows that the total merger fraction is dominated by binary mergers, and that the multiple merger fraction tends to be $1-2$ orders of magnitude smaller than the binary merger fraction. It also follows that $f_m/f_t\propto f_t^{2/3}$, namely the ratio between the multiple and total merger fractions, $f_m/f_t$, is higher where the total merger fraction $f_t$ itself is larger, for instance at higher redshift.
    \item The simulations theoretically predict that galaxy mergers in $\Lambda$CDM are more strongly clustered in time than a Poisson process. Our calculation of the ratio $f_m/f_t$ for a toy model where mergers occur following a Poisson process produces a multiple merger fraction of $f_m/f_t\approx 0.5f_t$, which is usually smaller than that measured empirically in the simulations. The difference between the empirical simulation results and the Poisson process toy model is even more pronounced when considering the full distribution of the time-to-closest-merger, which is much narrower in the simulation than a Poisson process would imply. Another toy model, in which mergers occur with fixed time intervals in between them, produces even (much) fewer multiple mergers and is hence significantly unrealistic.
    \item Further accentuating the notion of time-clustered mergers, we find that the second-closest merger in the evolutionary timeline of an arbitrary galaxy selected at an arbitrary cosmic time is typically within $\sim1$Gyr of the closest merger itself, even if that closest merger is several Gyr away from the selection time. That is, the occurrence of one merger is a strong predictor of another merger occurring around the same time. Relatedly, we find tentative evidence that four-way mergers are not as rare with respect to three-way mergers as three-way mergers are with respect to binary mergers. While four-way mergers are rather rare, these relative relations are further evidence of the non-random, clustered nature of galaxy mergers.
\end{itemize}

Quantifying the multiple merger fraction requires adopting some finite duration for the merger process. Otherwise, if mergers are considered instantaneous, it is only the merger rate, namely the number of mergers per unit time, that is well defined, but not the merger fraction, which is the fraction of galactic systems that are undergoing a merger at a given moment in time. In this work, we adopt a fiducial explicit form for the merger duration based on the literature; however, we seek results that are more general. This is because the duration of a merger, namely the length of the time window during which it is considered to be observable or ongoing in some other sense, may strongly depend on definitions and is generally not yet very well constrained or known. Although the quantitative values inferred for the merger fractions are closely tied to the quantitative choice for the merger duration, we have overcome this limitation by showing that the relation $f_m/f_t\approx 0.5f_t^{2/3}$ is quite general. That is, whatever galaxy sample is selected and whatever definition is adopted for what constitutes a merger and how long mergers last, our robust quantitative prediction is that the fraction of merging systems that are multiple rather than merely binary mergers is approximately equal to the fraction of all systems that are experiencing some kind of merger at all.

This means that, generally speaking, multiple mergers are subdominant within the total population of mergers. In what might initially appear as being in tension with this conclusion, \citet{Moster_2013}, who presented distributions for the time separation between adjacent halo mergers in a cosmological simulation, a quantity that is very similar to our $\Delta t_2-\Delta t_1$, found that at low redshift most adjacent mergers are separated by less than three halo dynamical times, a result that they interpreted as implying that mergers should not be treated as successive isolated events. Their results, however, are in good agreement with ours when it is considered that, in a large fraction of the parameter space (in particular towards lower mass and more major mergers), most galaxies do not have adjacent mergers to begin with, as they only experience a single merger throughout their formation history. For example, while Fig.~1 in \citet{Moster_2013} shows that of all low-mass halos that experience a merger more major than a 1:4 ratio at $z<1$, $\approx8\%$ will experience another such merger within three halo dynamical times and only $\lesssim3\%$ will experience another such merger with a longer time separation, it is still the case that $\sim90\%$ of such halos will not experience another such merger at all, thus making multiple mergers of this type a rather atypical occurrence.

This subtlety about the comparison to the results of \citet{Moster_2013} further highlights our finding that only a fraction $F$ of galaxies experience any merger at all throughout their evolution, a notion that plays a role in the comparison between our simulation results and our toy scenarios, in particular the Poisson process scenario. In our fiducial Poisson process scenario, we assume that galaxy mergers have an ergodic-like nature, that is, that the probability of experiencing a merger is evenly distributed in the galaxy population. When we abandon that assumption and instead assume that it is evenly distributed only amongst those galaxies that undergo any merger at all during their evolution, we obtain somewhat larger multiple merger fractions in the Poisson process scenario, even if the gap with the actual simulation results does not fully close. While the galaxy merger rate as traditionally defined (e.g.~in RG15, with which our novel method for calculating the binary merger fraction agrees very well) relies only on the merger count and is completely insensitive to how the mergers are distributed amongst different galaxies, studying the multiple merger fraction exposes the notion that not all galaxies (even those selected with a given mass and redshift) are equally likely to experience a merger; in other words, that the ensemble average of the merger rate per galaxy is not equal to the time average of the merger rate of individual galaxies in the ensemble.

The notion that galaxy mergers are more clustered in time than random arrivals in a Poisson process could have implications for several physical effects caused by or related to galaxy mergers. First, the dynamics of the mergers is likely to be affected by the complex multiple-body interaction and imprint signatures on the remnant properties. In addition, the concurrent infall of several satellites can affect the properties of their accretion histories and internal properties \citep{Trelles_2022}. It could similarly affect the evolution and outcome of mergers of supermassive black holes \citep{SayebM_24a} and of nuclear star clusters.

Although not the primary focus of this work, it is worth noting that we have also, for the first time, compared the total merger fraction between the original Illustris and the IllustrisTNG simulations and found some intriguing differences. In TNG, the mass dependence of the merger fraction is shallower at low masses and steeper at high masses, and the merger rate decreases towards later cosmic time less fast than in the original Illustris. A comparison of those trends with other models and observations is warranted and is reserved for future work.

We also leave to future work an application of our approach to halo mergers (rather than galaxy mergers as done here), which at the high-mass end would correspond to mergers of galaxy groups and clusters. Anecdotally, as discussed in the Introduction, many cluster mergers are reported to be comprised of a number of merging components, potentially corresponding to our definition of a multiple merger. Qualitatively, our results indeed predict higher multiple-to-total merger fractions at higher masses. A proper quantification of the halo multiple merger fraction and a quantitative comparison to observations would be interesting to pursue in the future.

\section{Data Availability} \label{sec:data_availability}

The code used to generate our results from this work is available in Zenodo \citep{jonathan_mack_2025_16748248}. Users interested in the latest version of the code or in any other versions should go to \url{https://doi.org/10.5281/zenodo.16748248}.

\section{Acknowledgments}
\label{sec:acknowledgments}

The Flatiron Institute is supported by the Simons Foundation. The manuscript was edited with the assistance of Writefull (\citealt{Writefull2024}).

\bibliography{mg23}

\end{document}